\documentclass[twoside,twocolumn,9pt]{article}
\usepackage{extsizes}
\usepackage[super,sort&compress,comma]{natbib} 
\usepackage[version=3]{mhchem}
\usepackage[left=1.5cm, right=1.5cm, top=1.785cm, bottom=2.0cm]{geometry}
\usepackage{balance}
\usepackage{mathptmx}
\usepackage{amsmath}
\usepackage{sectsty}
\usepackage{graphicx} 
\usepackage{lastpage}
\usepackage[format=plain,justification=justified,singlelinecheck=false,font={stretch=1.125,small,sf},labelfont=bf,labelsep=space]{caption}
\usepackage{float}
\usepackage{fancyhdr}
\usepackage{fnpos}
\usepackage[english]{babel}
\addto{\captionsenglish}{%
  
}
\usepackage{array}
\usepackage{droidsans}
\usepackage{charter}
\usepackage[T1]{fontenc}
\usepackage[usenames,dvipsnames]{xcolor}
\usepackage{setspace}
\usepackage[compact]{titlesec}
\usepackage{hyperref}
\usepackage{subfiles}
\usepackage{xcolor}

\usepackage{epstopdf}

\usepackage{graphicx}
\usepackage{color}
\usepackage{epstopdf, epsfig}
\usepackage{psfrag}
\usepackage{amsmath, bm}
\usepackage{amssymb}
\usepackage{enumitem}
\usepackage[toc,page]{appendix}
\usepackage{float}
\usepackage{tabu}
\usepackage{array}
\usepackage{subcaption}

\definecolor{cream}{RGB}{222,217,201}

\begin{document}

\pagestyle{fancy}
\thispagestyle{plain}
\fancypagestyle{plain}{
\renewcommand{\headrulewidth}{0pt}
}

\makeFNbottom
\makeatletter
\renewcommand\LARGE{\@setfontsize\LARGE{15pt}{17}}
\renewcommand\Large{\@setfontsize\Large{12pt}{14}}
\renewcommand\large{\@setfontsize\large{10pt}{12}}
\renewcommand\footnotesize{\@setfontsize\footnotesize{7pt}{10}}
\makeatother

\renewcommand{\thefootnote}{\fnsymbol{footnote}}
\renewcommand\footnoterule{\vspace*{1pt}%
\color{cream}\hrule width 3.5in height 0.4pt \color{black}\vspace*{5pt}} 
\setcounter{secnumdepth}{5}

\makeatletter 
\renewcommand\@biblabel[1]{#1}            
\renewcommand\@makefntext[1]%
{\noindent\makebox[0pt][r]{\@thefnmark\,}#1}
\makeatother 
\sectionfont{\sffamily\Large}
\subsectionfont{\normalsize}
\subsubsectionfont{\bf}
\setstretch{1.125} 
\setlength{\skip\footins}{0.8cm}
\setlength{\footnotesep}{0.25cm}
\setlength{\jot}{10pt}
\titlespacing*{\section}{0pt}{4pt}{4pt}
\titlespacing*{\subsection}{0pt}{15pt}{1pt}

\fancyfoot{}
\fancyfoot[LO,RE]{\vspace{-7.1pt}\includegraphics[height=9pt]{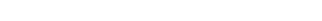}}
\fancyfoot[CO]{\vspace{-7.1pt}\hspace{13.2cm}\includegraphics{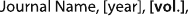}}
\fancyfoot[CE]{\vspace{-7.2pt}\hspace{-14.2cm}\includegraphics{head_foot/RF}}
\fancyfoot[RO]{\footnotesize{\sffamily{1--\pageref{LastPage} ~\textbar  \hspace{2pt}\thepage}}}
\fancyfoot[LE]{\footnotesize{\sffamily{\thepage~\textbar\hspace{3.45cm} 1--\pageref{LastPage}}}}
\fancyhead{}
\renewcommand{\headrulewidth}{0pt} 
\renewcommand{\footrulewidth}{0pt}
\setlength{\arrayrulewidth}{1pt}
\setlength{\columnsep}{6.5mm}
\setlength\bibsep{1pt}

\makeatletter 
\newlength{\figrulesep} 
\setlength{\figrulesep}{0.5\textfloatsep} 

\newcommand{\topfigrule}{\vspace*{-1pt}%
\noindent{\color{cream}\rule[-\figrulesep]{\columnwidth}{1.5pt}} }

\newcommand{\botfigrule}{\vspace*{-2pt}%
\noindent{\color{cream}\rule[\figrulesep]{\columnwidth}{1.5pt}} }

\newcommand{\dblfigrule}{\vspace*{-1pt}%
\noindent{\color{cream}\rule[-\figrulesep]{\textwidth}{1.5pt}} }

\makeatother

\twocolumn[
  \begin{@twocolumnfalse}
{\includegraphics[height=30pt]{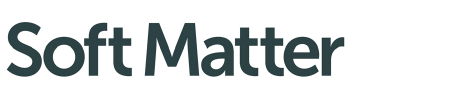}\hfill\raisebox{0pt}[0pt][0pt]{\includegraphics[height=55pt]{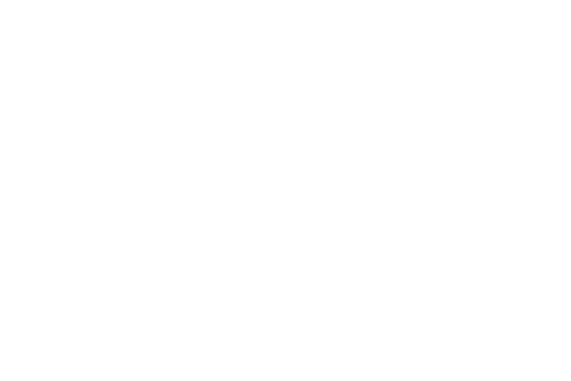}}\\[1ex]
\includegraphics[width=18.5cm]{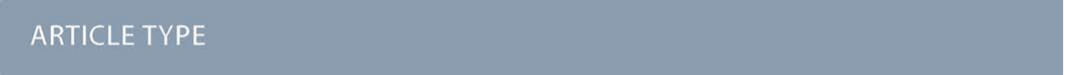}}\par
\vspace{1em}
\sffamily
\begin{tabular}{m{4.5cm} p{13.5cm} }

\includegraphics{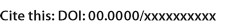} & \noindent\LARGE{\textbf{The fluid dynamics of a viscoelastic fluid dripping onto a substrate$^\dag$}} \\
\vspace{0.3cm} & \vspace{0.3cm} \\

& 
\noindent\large{Konstantinos Zinelis$^{\ast}$\textit{$^{a,b}$}, Thomas Abadie\textit{$^{c}$}, Gareth H. McKinley\textit{$^{b}$}}, and Omar K. Matar\textit{$^{a}$}\\

\includegraphics{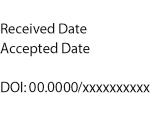} & \noindent\normalsize{Extensional flows of complex fluids play an important role in many industrial applications, such as spraying and atomisation, as well as microfluidic-based drop deposition. The Dripping-on-Substrate (DoS) technique is a conceptually-simple, but dynamically-complex, probe of the extensional rheology of low-viscosity, non-Newtonian fluids. It incorporates the capillary-driven thinning of a liquid bridge, produced by a single drop as it is slowly dispensed from a syringe pump onto a solid partially-wetting substrate. By following the filament thinning and pinch-off process the extensional viscosity and relaxation time of the sample can be determined. Importantly, DoS allows experimentalists to measure the extensional properties of lower viscosity solutions than is possible with commercially available capillary break-up extensional rheometers. Understanding the fluid mechanics behind the operation of DoS will enable users to optimise and extend the performance of this protocol. To achieve this understanding, we employ a computational rheology approach, using adaptively-refined time-dependent axisymmetric numerical simulations with the open-source Eulerian code, \textit{Basilisk}. The volume-of-fluid technique is used to capture the moving interface, and the log-conformation transformation enables a stable and accurate solution of the viscoelastic constitutive equation. Here, we focus on understanding the roles of surface tension, elasticity and finite chain extensibility in controlling the Elasto-Capillary (EC) regime, as well as the perturbative effects that gravity and substrate wettability play in setting the evolution of the self-similar thinning and pinch-off dynamics. To illustrate the interplay of these different forces, we construct a simple one-dimensional model that captures the initial rate of thinning when the dynamics are dominated by a balance between inertia and capillarity. This model also captures the structure of the transition region to the nonlinear EC regime in which the rapidly growing elastic stresses in the thread balance the capillary pressure as the filament thins towards breakup. Finally, we propose a fitting methodology based on the analytical solution for FENE-P fluids to improve the accuracy in determining the effective relaxation time of an unknown fluid.} \\

\end{tabular}

 \end{@twocolumnfalse} \vspace{0.6cm}

  ]

\renewcommand*\rmdefault{bch}\normalfont\upshape
\rmfamily
\section*{}
\vspace{-1cm}


\footnotetext{\textit{$^{a}$~Department of Chemical Engineering, Imperial College London, London SW7 2AZ, United Kingdom}}
\footnotetext{\textit{$^{b}$~Department of Mechanical Engineering, Massachusetts Institute of Technology, Cambridge, MA 02139, USA. }}
\footnotetext{\textit{$^{c}$~School of Chemical Engineering, University of Birmingham, Birmingham B15 2TT, United Kingdom}}



\section{Introduction}

The formation of liquid droplets results from flows featuring complex topological changes of a deformable fluid via the creation and breakup of filaments. Thread formation and droplet pinch-off are of central importance to atomisation and sprays  \citep{peregrine_1996,Villermaux2007, Eggers2008}, inkjet printing \citep{Basaran2013, Rosello2019, Lohse2022}, dripping \citep{Eggers1997, Basaran2002, Rajesh2022}, agrochemicals \citep{Makhnenko2021, Xu2021}, coatings \citep{Owens2011}, and also features in physiological flows such as sneezing and coughing \citep{Scharfman2016}. In a complex fluid, the extensional kinematics in the filaments arise from the streamwise velocity gradients and shear-free boundary conditions in the neck region as the breakup is approached. The extensional viscosity, $\eta_E$, which resists the thinning process is larger than its shear counterpart, $\eta_s$, ($\eta_{E} = 3 \eta_s$ for a Newtonian fluid \citep{Dinic2022}), and the dynamics of the drop formation process are determined by the combined effects of capillary, inertial, and viscous forces. The addition of a high molecular weight flexible polymer such as poly(ethylene oxide) (PEO) to a Newtonian solvent like water or water-glycerol mixtures, has been observed experimentally to strongly influence the neck evolution and thinning dynamics, delaying the final pinch-off \citep{Eggers1997, McKinley2005, Eggers2014, Deblais2020}. For dilute polymer solutions, the fluid viscoelasticity can lead to a large increase of the extensional viscosity (in comparison to the shear viscosity) by a factor of $10^2  - 10^3$. It is therefore important to measure accurately parameters such as $\eta_E$ and the polymer relaxation time $\tau$, which characterise the extensional rheology of the solution as they influence the filament thinning process and the timescales controlling droplet formation in complex fluids \citep{Robertson2022}.

\begin{figure*}[h!]
\centering
  \includegraphics[width=\textwidth]{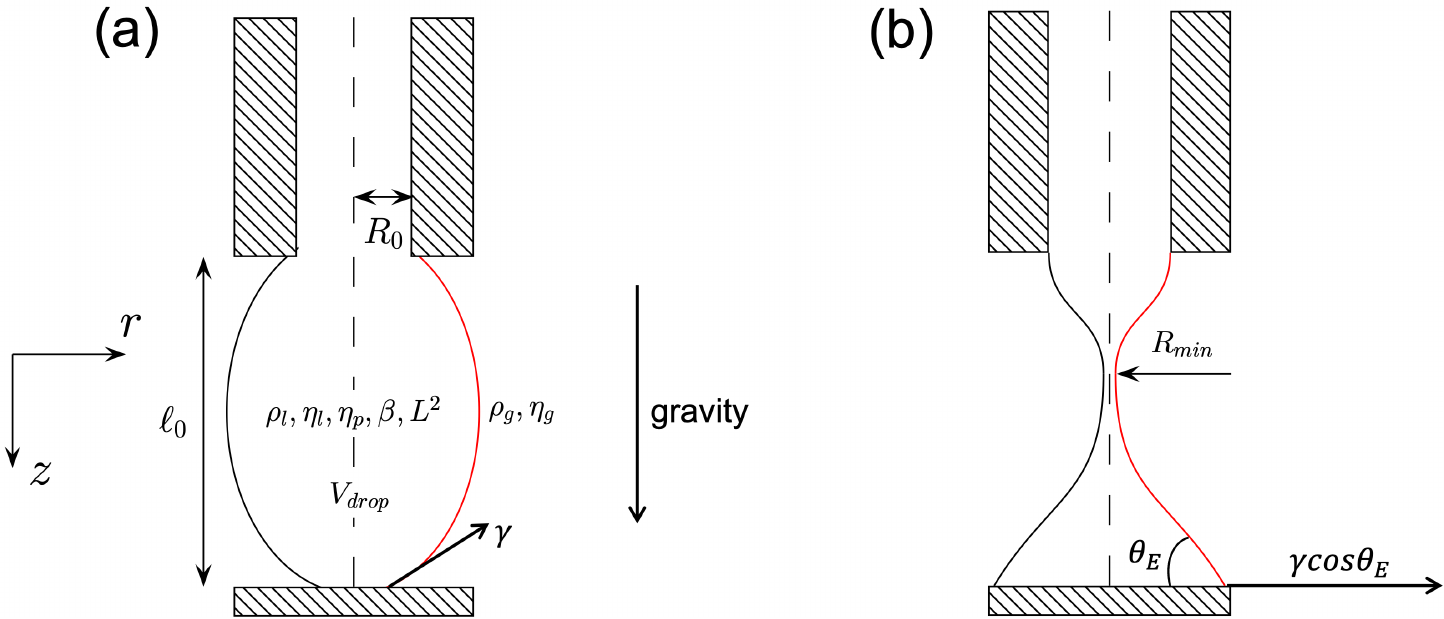}
  \caption{Simulation setup and interface evolution during two distinct stages: (a) initially, gravity-driven axial drop elongation followed by, (b) spreading over the solid substrate of prescribed wettability set by the equilibrium contact angle $\theta_E$. The simulation domain is indicated in red.}
  \label{fig:initialisation}
\end{figure*}
Various bespoke extensional rheometers have been developed based on a variety of configurations such as jetting, e.g. the Rayleigh Ohnesorge jetting extensional rheometer (ROJER) device \citep{Keshavarz2015,rojer2017}, spinning, flow through T-junctions and cross-slot configurations (e.g. OSCER), as well as boundary separation techniques such as the Filament Stretching Extensional Rheometer (FiSER) and the Capillary Breakup Extensional Rheometer (CaBER) \citep{Bazileveskii1981, Yesilata2006, Keshavarz2015, Sharma2015, Dinic2022}. The CaBER rheometer design exploits the capillarity-driven thinning of a stretched fluid bridge in the absence of imposed external forces and is suitable for studying a range of mobile complex fluid samples (i.e. low and moderate viscosity) over a range of Hencky strains. However, as the viscosity of the test fluid is lowered towards that of water, inertial effects become increasingly important and limit successful operation \citep{Rodd2005}.

The recently developed Dripping-onto-Substrate (DoS) protocol \citep{Dinic2015, Dinic2017a, Dinic2019}, which involves the slow dripping of a single drop through a narrow needle (needle radius $R_0 \sim 1$ mm) onto a partially-wetting substrate, as illustrated in Figure \ref{fig:initialisation}(a), is suitable for extensional rheology measurements of $O(\mu L)$ volumes of complex fluids, enabling tests on expensive/scarce material samples, such as protein solutions \citep{Lauser2021} whilst mitigating undesirable shear and inertial effects. The DoS rheometry technique also facilitates the measurement of the extensional properties of lower-surface tension and lower-viscosity fluids at higher attainable extension rates than commercially available capillary breakup instrumentation (e.g. the CaBER device) \citep{Dinic2022}.
Effective DoS measurements have been reported in studies of the extensional rheology for various polymer solutions, suspensions, inks, viscoelastic surfactant fluids, cosmetics and food materials, as well as protein solutions and associative polysaccharide systems \citep{Dinic2017a, Dinic2017b, Suteria2019, Walter2019, Omidvar2019, Rosello2019, Rassolov2020, Jimenez2020, MartinezNarvaez2021, Lauser2021, Dinic2022}. DoS protocols have helped to elucidate the non-Newtonian response to extensional deformations for fluids which do not exhibit any measurable viscoelastic behaviour in conventional shear rheometer or extensional rheometer experiments. By understanding the sequence of local balances among the capillary, inertial, viscous, and elastic forces acting on the thinning thread attached to the drop as it spreads laterally on the substrate \citep{Dinic2022}, the DoS process can be shown to be characterised by the sequential emergence of a number of distinguished regimes and distinct neck shapes. 

At early thinning times, and for low-viscosity Newtonian samples, an Inertio-Capillary (IC) response described by power-law dynamics dominates the thinning process leading to a conically-shaped neck, close to $R_{min}(t)$, as indicated in Figure \ref{fig:initialisation}(b). In contrast, in fluids of higher viscosity, a distinct Visco-Capillary (VC) regime is established in which the neck is slender and almost cylindrical in shape and the radius $R_{min}(t)$ decreases linearly in time. For viscoelastic fluids, an additional Elasto-Capillary (EC) thinning regime is observed as thread pinch-off is approached. This is characterised by an exponential decrease in $R_{min}(t)$ with time and the formation of a thin and axially uniform cylindrical thread connecting the deposited drop to the residual fluid that is pinned by the nozzle. In the final stages of the breakup process, this EC regime gives way to a Terminal Visco-Elasto-Capillary (TVEC) region in which the radius once again decreases to zero linearly in time. This terminal regime is controlled by the finite extensibility of the polymer chains that are dissolved in the low-viscosity solvent.
In this work, we use time-dependent, free-surface numerical simulations of a canonical constitutive model for polymer solutions (the FENE-P model \citep{Bird1987}) to study filament deformation and the breakup dynamics that underpin DoS rheometry. We extend previous experimental work examining the role of substrate wetting and gravitational forces for wormlike micellar solutions \citep{Wu2020} and inertio-capillary dynamics \citep{Lauser2021} to account for the effects of nonlinear viscoelasticity and finite chain extensibility that are captured by the FENE-P constitutive equation. The thinning fluid thread and the substrate wettability are parameterised by the Ohnesorge, Deborah, and Bond numbers, as well as a macroscopic contact angle, respectively. Our computational rheology experiments help identify the existence of optimal parameter ranges in which experimental inaccuracies for these small-scale, high-speed rheological measurements are minimised.  
The rest of this article is organised as follows. In Section \ref{sec:Formulation}, the problem formulation and the numerical framework used for performing the simulations are presented. We then investigate how the dynamics of DoS rheometry are affected by substrate wettability and gravitational forces, and representative results are discussed in Section \ref{sec:Results}. In Section \ref{sec:Fitting}, we use a simple one-dimensional model to capture the main dynamical features identified by our full numerical simulations and explore how the data obtained from systematic parametric variations of the DoS rheometry protocol can improve the analysis of extensional rheology measurements with unknown samples. Finally, recommendations for better practices in DoS rheometry and concluding remarks are presented in Sections \ref{sec:Fitting} and \ref{sec:Conclusions}, respectively.
%

\section{Formulation and Methodology \label{sec:Formulation}}

\subsection{Flow configuration}

Figure \ref{fig:initialisation} shows the configuration of a typical DoS experiment. We use an axisymmetric description of the filament shape $R(z,t)$ as a prolate pendant drop of length $\ell_0$ is brought into contact with a partially-wetting substrate. The fluid is an incompressible, viscoelastic solution containing polymer chains of finite extensibility $L^2$ which issues from a nozzle of length $\ell_{nozzle}$ and initial radius $R_{0}$. The fluid of density $\rho_l$ and dynamic viscosity $\eta_l=\eta_p+\eta_s$ is surrounded by a gas of density $\rho_g$ and dynamic viscosity $\eta_g$; here, $\eta_p$ and $\eta_s$ denote the polymer and solvent contributions to the total viscosity, respectively. Additionally, we use $\beta=\eta_s / \eta_l$ to express the relative contribution of the solvent viscosity $\eta_s$ to the total dynamic viscosity of the polymer solution $\eta_l$. The droplet issuing from the nozzle is brought into contact with a solid substrate located at a distance $\ell_0$ below the nozzle exit resulting in a characteristic aspect ratio of $\ell_0 / 2 R_0$. The substrate wettability is parameterised by a macroscopic equilibrium contact angle $\theta_{E}$. 

The initial volume of the droplet $V_{drop}$ is ensured to be larger than the corresponding volume of a fluid column of radius $R_0$ and length $\ell_0$ for stability considerations \citep{McKinley2005}. As shown schematically in Figure \ref{fig:initialisation}(a), the pendant drop develops a prolate amphora shape whose subsequent spreading across the substrate is influenced by gravitational forces, wettability effects, as well as its rheological response. The lateral spreading of the droplet foot and conservation of volume leads to the development of a thin polymeric thread of characteristic radius $R_{min}(t)$, which connects the two hemispherical droplets that are attached to the nozzle and spreading across the substrate, respectively, as illustrated in Figure \ref{fig:initialisation}(b). The thinning and eventual break up of this thread is influenced by a delicate balance of capillary, inertial, viscous, and elastic forces, as will be discussed below. Understanding this evolving balance is what enables the DoS configuration to be employed in extensional rheometry.

\subsection{Governing equations and numerical method\label{sec:Equations}}
In what follows we non-dimensionalise the governing equations by scaling with the characteristic length $R_0$, an inertio-capillary velocity $U_R= \sqrt{\gamma / \left( \rho_l R_0 \right)}$, and the Rayleigh time $t_R = R_0/U_R = \sqrt{\rho_l R_0^3/\gamma}$. All stresses and the isotropic pressure thus can be made dimensionless using a pressure scale $\rho_{l} U_R^2 = \gamma/R_0$.
The filament thinning dynamics are then governed by the following dimensionless continuity and momentum conservation equations: 
\vspace{-0.2in}
\begin{center}
\begin{equation}
\boldsymbol{\tilde{\nabla} \cdot \mathbf{\tilde{u}}} = 0,
\label{eq:scaled-continuity}
\end{equation}
\begin{equation}
\Tilde{\rho} \left(\dfrac{\partial \mathbf{\Tilde{u}}}{\partial \Tilde{t}} +  \mathbf{\Tilde{u}} \cdot \Tilde{\boldsymbol{\nabla}} \mathbf{\Tilde{u}} \right)  = - \Tilde{\boldsymbol{\nabla}} \Tilde{p} + Oh\left( \beta \Tilde{\boldsymbol{\nabla}} \cdot \Tilde{\boldsymbol{\sigma}}_s + \frac{\left(1-\beta \right)}{De} \Tilde{\boldsymbol{\nabla}} \cdot \Tilde{\boldsymbol{\sigma}}_p \right) +\Tilde{\kappa} \Tilde{\delta} \mathbf{n} + Bo \mathbf{\Tilde{g}},
\label{eq:scaled_NS}
\end{equation} 
\end{center}
where the tildes designate dimensionless variables; $\tilde{t}$, $\tilde{\rho}$, $\mathbf{\tilde{u}}$, $\tilde{p}$, $\Tilde{\boldsymbol{\sigma}}_s$, $\Tilde{\boldsymbol{\sigma}}_p$, $\tilde{\kappa}$, $\tilde{\delta}$, $\mathbf{n}$, and $\mathbf{\Tilde{g}}$
correspond to dimensionless time, density, velocity, pressure, solvent and polymeric stress components, interfacial curvature, the Dirac delta function, the outward-pointing unit normal vector to the interface, and the gravitational acceleration, respectively. 
A one-fluid Volume-of-Fluid (VOF) approach \citep{Popinet2009} is used, where the volume fraction $c$ of the liquid phase is advected in every computational cell as $\partial{c}/\partial \tilde{t}+\tilde{\mathbf{u}}\cdot \Tilde{\boldsymbol{\nabla}} c=0$. Additionally, 
the local density $\Tilde{\rho}$ and viscosity $\Tilde{\eta}$ are determined by:
 \begin{eqnarray}
    \Tilde{\rho} & = & c + (1-c)\dfrac{\rho_{g}}{\rho_l},\\
    \Tilde{\eta} & = & c + (1-c) \dfrac{\eta_{g}}{\eta_l}.
    \label{eq:properties}
\end{eqnarray}
In Eq. (\ref{eq:scaled_NS}), $De=\tau/(R_0/U_R)= \tau / t_R$ represents an intrinsic Deborah number, which captures the ratio of the relaxation time of the polymer $\tau$ to the Rayleigh flow time scale $t_R$; $Oh=\eta_l/\sqrt{\rho_l\gamma R_0}$ is the Ohnesorge number that represents the relevant contribution of capillary and viscous forces, and $Bo = \rho_l g R_0 \ell_0 / \gamma$ is a Bond number that characterises the balance between gravity and capillarity in the initial pendant drop as it touches the substrate.
The dimensional viscous stress tensor arising from the solvent is given by $\boldsymbol{\sigma_{s}}=\eta_{s}(\boldsymbol{\nabla} \mathbf{u}+( \boldsymbol{\nabla}\mathbf{u})^{T})$, and $\boldsymbol{\sigma_{p}}$ is the dimensional polymeric stress contribution to the total stress given here by the FENE-P constitutive equation:
\vspace{-0.2in}
        \begin{center}
        \begin{equation}
        \boldsymbol{\sigma_{p}}=\frac{\eta_{p}} {\tau} \left( \frac{\mathbf{A}}{1-\frac{\operatorname{tr}(\mathbf{A})} {L^2}} 
        -\mathbf{I} \right),
        \label{eq:stress}
        \end{equation}
        \end{center}
where $L^2$ is the polymer chain finite extensibility, and $\mathbf{A}$ is the dimensionless chain conformation tensor whose transport is governed by the following dimensionless equation: 
\begin{equation}
 \frac{\partial \mathbf{A}} {\partial \Tilde{t}} + \mathbf{\Tilde{u}} \cdot \boldsymbol{\Tilde{\nabla}} \mathbf{A} -\left(\boldsymbol{\Tilde{\nabla}} \mathbf{\Tilde{u}} \cdot \mathbf{A}+ \mathbf{A} \cdot \boldsymbol{\Tilde{\nabla}} \mathbf{\Tilde{u}}^{T}\right)= -\frac{1}{De} \left(\frac{\mathbf{A}}{1-\frac{\operatorname{tr}(\mathbf{A})}{L^2}}-\mathbf{I} \right).
 \label{eq:scaled-conformation}        
 \end{equation}

We avoid numerical instabilities due to rapid growth of elastic stresses during the formation of the elastic thread (High-Weissenberg Number Problem \citep{Renardy2000}) via the log-conformation transformation \citep{Fattal2005}. The local time-varying Weissenberg number $Wi$ is the  local strain rate normalised by the polymer relaxation time \citep{Tirtaatmadja2006}:
\begin{equation}
    Wi =  \tau \dot{\epsilon}_{min} = -2 \tau  \ \dfrac{ d \left(\log(R_{min})\right)}{dt} = -2 De  \ \dfrac{ d \left(\log(\Tilde{R}_{min})\right)}{d\Tilde{t}}.
    \label{eq:Wi}
\end{equation}
We use this dimensionless strain rate to assess the thinning rate of the viscoelastic filament. The tildes, which designate dimensionless variables, are suppressed henceforth.
%
%
\begin{figure*}[ht!]
\centering
  \includegraphics[height=9cm]{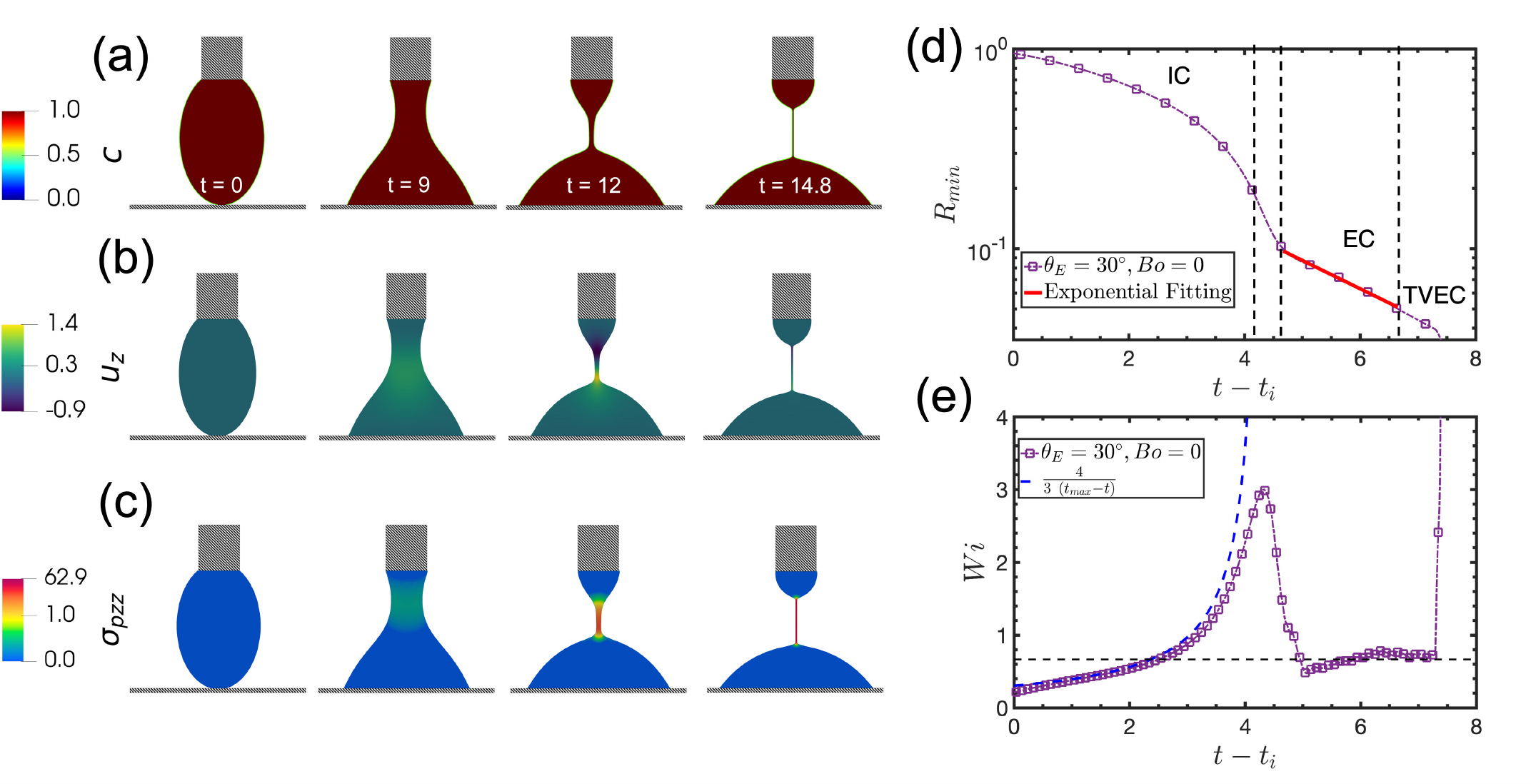}
  \caption{Contour plots of (a) the volume fraction, (b) the axial velocity component, and (c) the axial polymeric stress at different times which correspond to the initialisation, inertio-capillary thinning regime, the onset of the elasto-capillary regime, and the terminal thinning regime of the viscoelastic filament (which is controlled by the polymer finite extensibility). The formation of the characteristic thin viscoelastic thread between two beads is highlighted at $t=14.8$. (d) Representation of the evolution of the minimum filament $R_{min}$ in time, shifted by $t_i$ which is the time when the influence of the initial condition ceases (here $t_i = 7.46$); The simulations capture the three main characteristic regimes of filament thinning in DoS rheometry: i.e. Inertio-Capillary (IC), Elasto-Capillary (EC), and Terminal Visco-Elasto-Capillary (TVEC) regimes. The vertical dashed lines indicate the end of the IC regime and the subsequent transition to the exponential thinning, the onset of the EC regime, and finally the beginning of the TVEC thinning regime, respectively. (e) Temporal evolution of the local dimensionless strain rate, $Wi$, which shows good agreement between the numerical simulations and the theoretical prediction for the initial inertio-capillary dynamics (blue dashed line), the constant $Wi$ plateau which corresponds to the expected thinning-rate ($\sim - 1 / (3 De)$ \citep{entov_yarin_1984, Bazileveskii1990, Entov1997, Tirtaatmadja2006}), during the exponential thinning, and the rapid terminal divergence in $Wi$ when the polymer finite extensibility limit has been reached \citep{Entov1997, McKinley2005, Wagner2005}. Here, $t_{max}-t_i = 4.29$ (with $t_{max} = 11.75$ and $t_i=7.46$) corresponds to the time when the local $Wi$ attains its maximum value; this, in turn, corresponds to the end of the IC regime and the subsequent transition to the EC regime. The parameter values are listed in Table \ref{table_DoS}. For the corresponding movie please see the Electronic Supporting Information.}
  \label{fig:results_SetUp}
\end{figure*}
\subsection{Numerical set up\label{sec:num-setup}}
The simulations are performed with the open-source code \textit{Basilisk} \citep{Turkoz2018, Lopez-Herrera2019, Turkoz2021, Liu2023, zinelis2023transition}. The interface is reconstructed with a piecewise linear interface calculation (PLIC) technique \citep{Popinet2009, Lopez-Herrera2019}, and the height-function method is used to calculate the geometrical properties of the interface; this ensures accurate modelling of the capillarity-driven thinning of the polymeric solution \citep{Popinet2009, Popinet2018}. 
The simulation domain is a square of dimensions $1.67\pi R_0 \times 1.67\pi R_0$, which results in an aspect ratio $\approx 3$. The left boundary is the axial symmetry axis, while no-penetration Dirichlet conditions are considered for all the velocity components at the upper and lower boundaries, where a nozzle and solid substrate are considered, respectively. In addition, we follow the implementation by \citet{Lopez-Herrera2019} for the boundary conditions for all the polymeric stress tensor components. 

The contact line at the upper plane is pinned at $r=R_{0}$ to account for the nozzle exit, following the approach applied by \citet{Sakakeeny2021} so that drop oscillations can be effectively handled. The wetting of the solid substrate at the lower plane is modelled with a macroscopic contact angle boundary condition in combination with the height-function method \citep{Afkhami2008}. Although a no-slip boundary condition is imposed on the substrate, the velocity field for the interface advection is located at the centre of the cell faces and therefore results in an implicit slip condition at the contact line with a slip length which scales with the minimum grid size $\Delta r_{minimum}/2~$ \citep{Afkhami2008, Snoeijer2013}. Therefore, it is critical to specify small enough grid cell sizes to ensure that the macroscopic dynamics remain unaffected by the microscopic contact line dynamics (further details of the mesh convergence study are provided in the Electronic Supporting Information). Additionally, for the initial conditions of the DoS process at $t=0$, which is set when the polymeric droplet first touches the solid substrate, we consider that $r=R_{0}$ corresponds to the nozzle radius in the upper panel, and the polymeric chains are considered to be initially at rest which corresponds to $A_{zz}=A_{rr}=1$ everywhere within the liquid phase. The initial shape of an elongated polymeric droplet between the nozzle exit and the solid substrate is given by an elliptical equation:
\begin{equation}
 \left(\dfrac{z-2.45}{2.83}\right)^{2}+\left(\dfrac{r}{2}\right)^{2} -1 = 0 .
 \label{eq:droplet_shape}        
 \end{equation}

The grid cells in the simulation domain are refined according to a quadtree adaptive mesh refinement (AMR) scheme available in \textit{Basilisk} \citep{Popinet2009} based on the location of the interface, as well as on the regions where the gradients of the axial elastic stress are large. This allows for sufficient resolution of the extremely thin polymeric thread, which is formed between the two primary hemispherical beads (as also observed in experiments \citep{Dinic2015}). The adaptive mesh refinement enables accurate simulation of the thinning dynamics as breakup of the filament ($R_{min}(t) \rightarrow 0 $) is approached. Starting from a base grid resolution of $512 \times 512$ square cells for the whole domain, the adaptive scheme refines the cells down to a minimum square cell of size $\Delta r_{minimum}=0.0026$. 
%

\begin{figure*}[ht!]
\centering
   \includegraphics[width=0.9\textwidth]{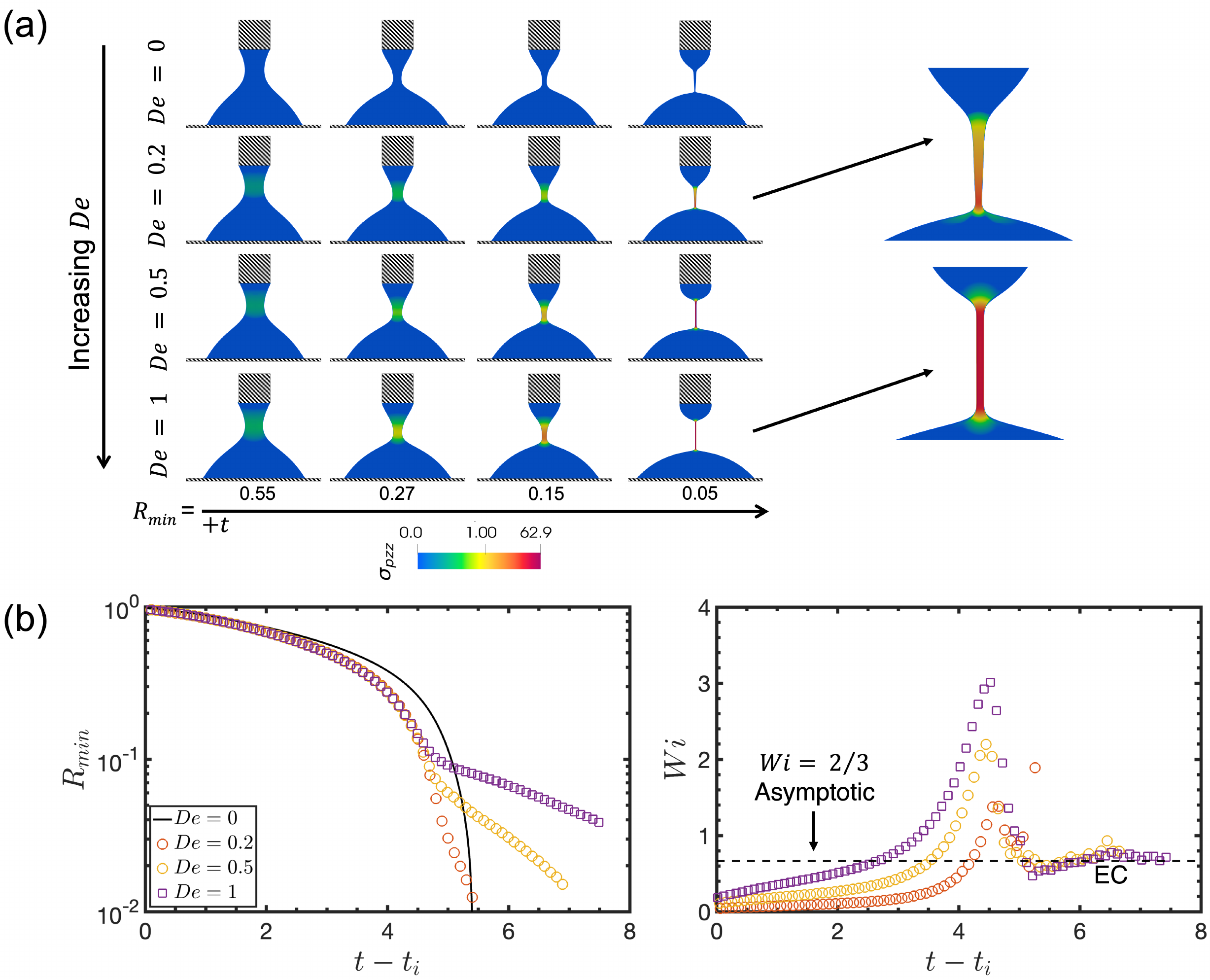}
  \caption{(a) Contour plots of the axial elastic stress component for different Deborah numbers, $De=0$, 0.2, 0.5 and 1, at dimensionless times that characterise different minimum radii of the filament, $R_{min} = 0.55$, 0.27, 0.15 and 0.05. Also shown are enlarged profiles of the viscoelastic stress distribution in the thread at $R_{min} = 0.05$ for $De=0.2$ and $1$, highlighting the effect of elasticity and the axially-uniform distribution of the polymeric stress in the thread for $De=1$. (b) Temporal profiles of the dimensionless minimum filament radius and the local dimensionless strain rate  $Wi=\tau \dot{\epsilon}$ for the same range of Deborah numbers, highlighting the emergence of a distinct Elasto-Capillary (EC) thinning regime at $De \gtrsim 0.5$ in which a constant dimensionless strain-rate $Wi = 2 /3$ is expected. The rest of the parameters remain unchanged from Table \ref{table_DoS}. For the corresponding movie please see the Electronic Supporting Information.}
  \label{fig:results_De}
\end{figure*}

\section{Results and Discussion \label{sec:Results}}
In this section, we provide a discussion of our results obtained from the numerical simulations of Dripping-Onto-Substrate (DoS) rheometry. Specifically, the effects of the parameters $De$, $L^2$, $Bo$, and $\theta_{E}$ on the thinning dynamics are analysed, and connections with the effectiveness of DoS rheometers are also established. 
\subsection{Numerical results}

\begin{figure*}[ht!]
    \includegraphics[width=\textwidth]{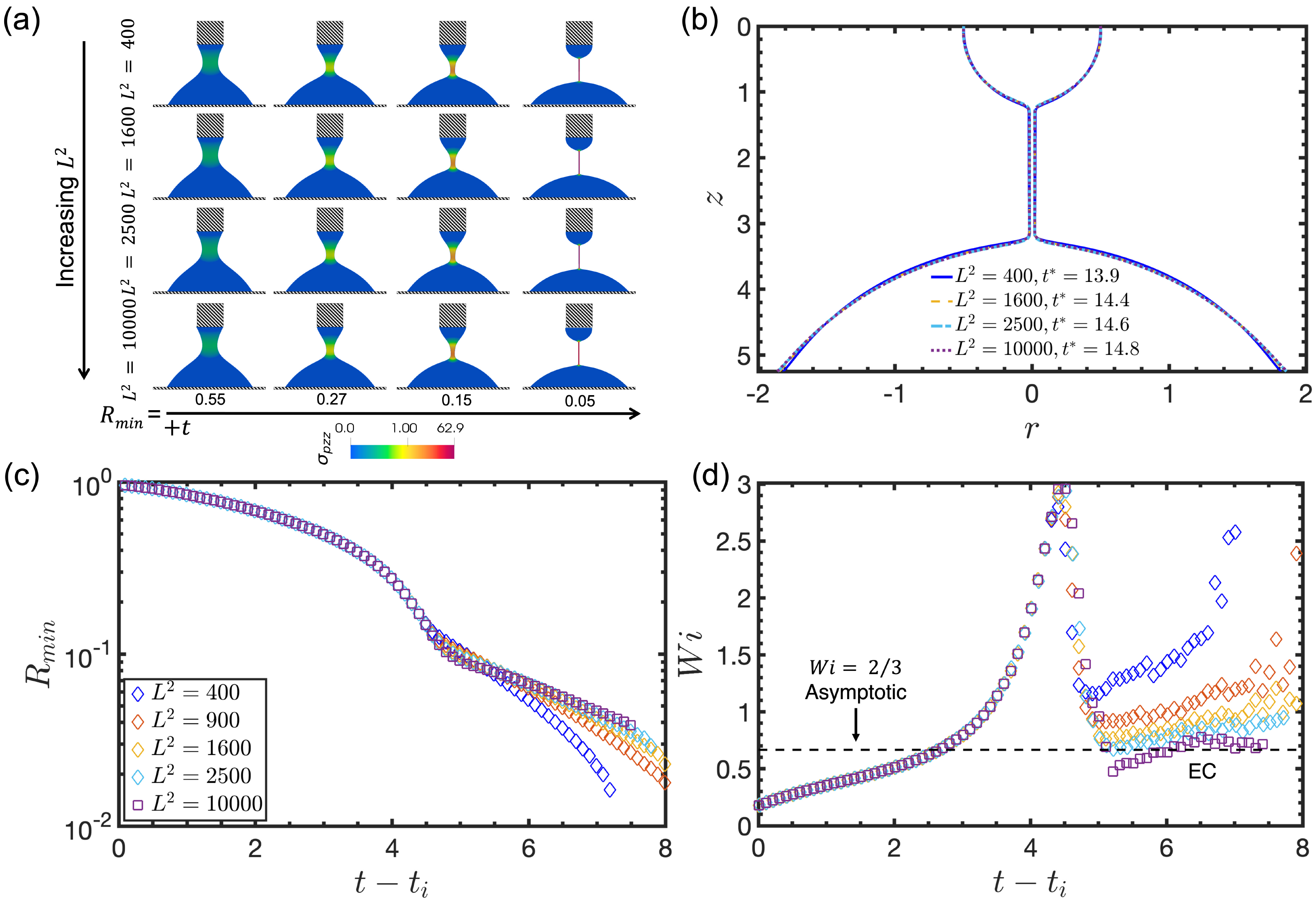}
  \caption{(a) Contour plots of the axial stress distribution in the viscoelastic thread at $R_{min} = 0.55$, 0.27, 0.15, and 0.05, for four representative polymer finite extensibility values $L^2 = 400$, 1600, 2500, and 10000 at fixed Deborah number ($De=1$). (b) Profiles of the viscoelastic filament at time $t^*$ when its minimum radius reaches a value of $5 \%$ of the nozzle radius. (c) Temporal evolution of the minimum filament radius $R_{min}(t)$, and (d) evolution in the local strain rate within the neck for $L^2 = 400$, 900, 1600, 2500, and 10000. The rest of the parameters remain unchanged from Table \ref{table_DoS}.}
  \label{fig:results_L2}
\end{figure*}

%
\begin{center}
\begin{table}[htb!]
\centering
\caption{Simulation parameters of DoS rheometry. Gravitational effects are initially neglected ($Bo=0$). The density and viscosity ratios $\rho_{g}/\rho_{l}$ and $\eta_{g}$/$\eta_{l}$ are the same as those used by \citet{zinelis2023transition}.}
\begin{tabular}{c c c c c c c c c }
\hline
$De$ & $Oh$ & $\beta$ & $L^2$ & $Bo$ & $\theta_{E}$ & $\rho_{g}/\rho_{l}$ & $\eta_{g}$/$\eta_{l}$\\  [0.5ex]
\hline
1 & 0.2 & 0.85 & 10000 & 0 & $30^\circ$ & 0.01 & 0.01\\
\end{tabular}
\label{table_DoS}
\end{table}
\end{center}
\vspace{-0.2in}
Figure \ref{fig:results_SetUp} shows the simulation results for a weakly viscoelastic fluid characterised by $De=1$ and large finite polymer chain extensibility $ L^2 = 10000$, with additional material properties as defined in Table \ref{table_DoS}, where the density and viscosity ratios are the same as in \citet{zinelis2023transition}. Figure \ref{fig:results_SetUp}(\textit{a})-(\textit{c}) presents the evolution of the interfacial shape along with contour plots of the volume-fraction, axial velocity component, and axial polymeric stress, respectively, coloured by the corresponding magnitude of each quantity. These figures demonstrate the evolution of a DoS experiment from the time when the droplet first touches the solid substrate ($t=0$) until the establishment of a fully-developed viscoelastic thread ($t=14.8$) when $R_{min}$ reaches the minimum resolvable value in our simulation ($R_{min} \approx 0.05$). The snapshots are taken at times which correspond to the initialisation of the numerical simulation ($t=0$), to the subsequent establishment of the inertio-capillary (IC) ($t=9$) and elasto-capillary (EC) ($t=12$) regimes, and finally to the ultimate thinning regime when the finite extensibility limit of the polymeric chains is reached ($t=14.8$). 

The contours of axial velocity presented in Figure \ref{fig:results_SetUp}(\textit{b}) show the initial slow drainage of the filament which eventually leads to the final rapid breakup of the viscoelastic thread and the subsequent formation of two hemispherical beads (of different volumes) attached to the nozzle and the lower substrate. The contour plots of the axial polymeric component in Figure \ref{fig:results_SetUp}(\textit{c}) reveal the importance of elastic stresses during the formation of the polymeric thread. In particular, at early times, the axial polymeric stresses are weak, and inertial and capillary forces drive the filament thinning process. However, at intermediate times ($t \geq 12$), the axial polymeric stress increases rapidly as the filament radius continues to thin. This large elastic stress acts to stabilise the thinning thread and retard its eventual breakup. 

The IC, EC, and TVEC regimes are all indicated in Figure \ref{fig:results_SetUp}(\textit{d}), which uses a semi-log representation of the evolution in the minimum filament radius $R_{min}(t)$; here, the dimensionless time scale on the abscissa has been shifted by an initial offset, $t_i$ which denotes the time at which the residual effects of the initial spreading process become negligible, i.e., when $R_{min} \leq 1$ and the initial prolate drop has established a concave necked configuration. For these specific parameter conditions, we determine $t_i=7.46$. In Figure \ref{fig:results_SetUp}(\textit{d}), the development of the exponential elasto-capillary thinning process is also evident, as expected \citep{Dinic2015}, until the finite extensibility of the polymer chain (set by the magnitude of the parameter $L^2$) has been reached. 

To examine the accuracy with which the simulation describes the thinning of the polymeric filament, we also monitor the temporal evolution of the local Weissenberg number $Wi$ in Figure \ref{fig:results_SetUp}(\textit{e}) which evolves in a non-monotonic way \cite{Tirtaatmadja2006, Rajesh2022}. Inspection of this profile confirms the fact that the simulations capture accurately the IC regime during the early times of the thinning process along with the subsequent exponential decrease of the filament radius in the EC regime. This is characterised by the expected dimensionless thinning rate of $-2 \dot{R}_{min} / R_{min}=2 / (3De)$ which corresponds to a constant local Weissenberg number of $Wi = 2/3$. The final steep increase of $Wi$ at the very end of the simulation indicates that the polymeric chains become fully extended and the filament approaches the ``pinch-off" interfacial singularity. Accurate resolution of this TVEC regime requires very high levels of mesh refinement \citep{zinelis2023transition}, but is not essential for successfully identifying and utilising the exponential EC regime. We thus cease our simulations when the minimum thread radius falls below $R_{min} \leq 0.05$ and the strain rate starts to rapidly increase again. 
We proceed now with studying the competing roles of elasticity, finite extensibility, gravity, and substrate wettability, in designing robust and efficient DoS rheological measurements. 
\begin{figure*}[ht!]
 \includegraphics[width=\textwidth]{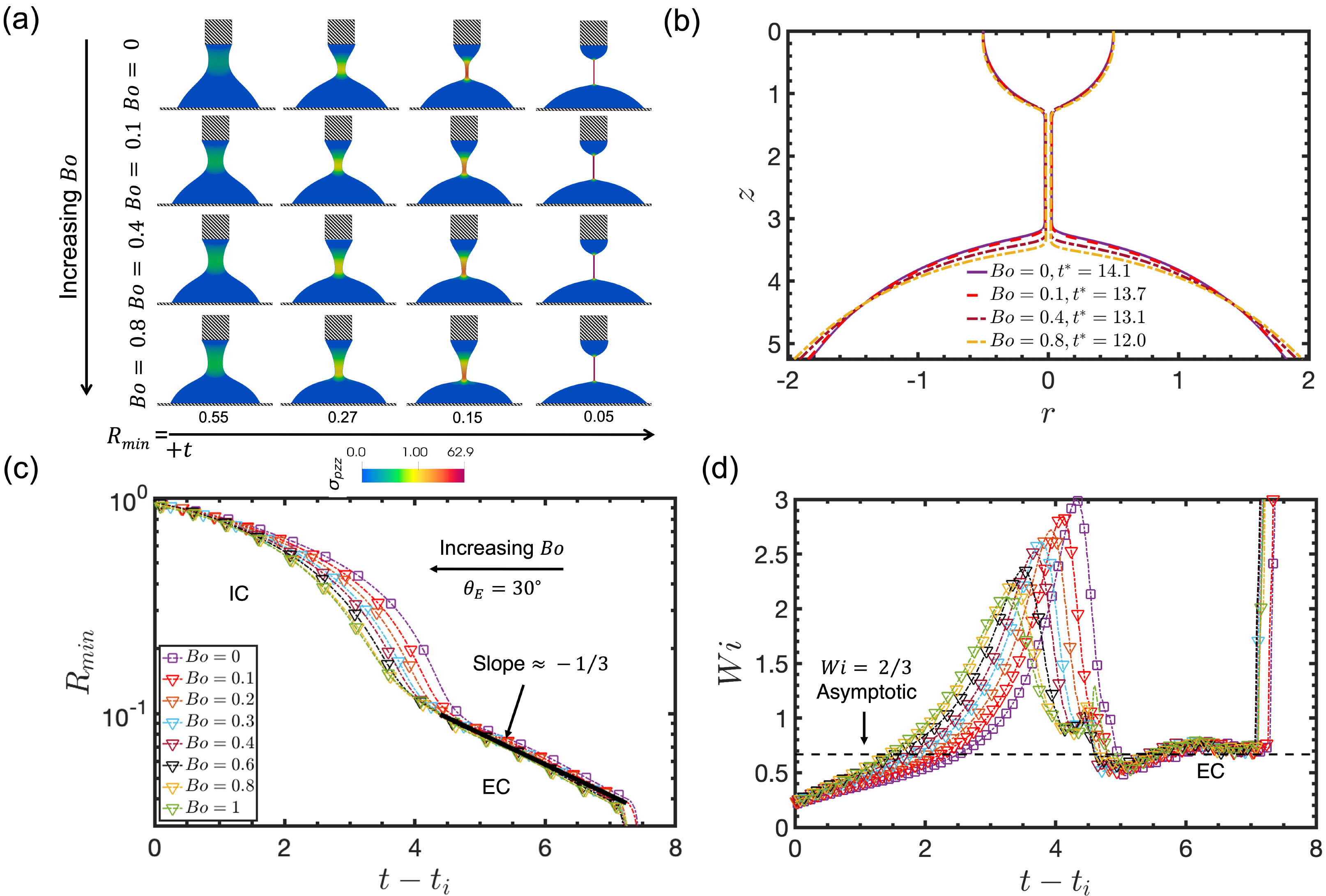}
  \caption{(a) Contour plots of the axial polymeric stress component at $R_{min} = 0.55$, 0.27, 0.15 and 0.05 for different $Bo=0$, 0.1, 0.4 and 0.8, highlighting the influence of the gravitational forces in the initial and terminal thinning dynamics. (b) Filament shape at $R_{min}=0.05$ for the same set of Bond numbers. Temporal profiles of (c) the dimensionless filament radius $R_{min}$ and (d) the local dimensionless strain rate $Wi$, highlighting the self-similar exponential thinning during the Elasto-Capillary (EC) regime for a range of different Bond numbers ($0 \leq Bo \leq 1$). The rest of the parameters are the same as given in Table \ref{table_DoS}. For the corresponding movie please see the Electronic Supporting Information.}
  \label{fig:Bo_results}
\end{figure*}

In Figure \ref{fig:results_De} we explore the level of elasticity in the polymeric solution above which elastic effects are detectable in DoS rheometers. Specifically, in Figure \ref{fig:results_De}(a), we show contour plots and interface profiles for $De=0$, 0.2, 0.5, and 1, respectively. It can be observed that for $De=0$ and 0.2, an approximately conical neck is formed, whilst, for $De\geq 0.5$, the thread becomes thin and of uniform thickness as also highlighted at the right of panel (a). In Figure \ref{fig:results_De}(b), at the left we show the exponential thinning of the filament radius $R_{min}(t)$ for $De=0, 0.2, 0.5$ and 1. A steeper slope of the filament thinning is observed with an increasing Deborah number, as expected according to the Oldroyd-B predictions ($\sim - 1 / (3 De)$ \citep{entov_yarin_1984, Bazileveskii1990, Entov1997, Tirtaatmadja2006}). For this reason, according to Eq. (\ref{eq:Wi}) the dimensionless strain rate  $Wi(t)$ written as a function of dimensionless parameters  ($Wi(t) = -2De \ d\left(\log(R_{min}(t))\right)/dt = 2 De /(3De) = 2/3$), is independent of the Deborah number during the EC regime. Hence, we expect a plateau in the dimensionless strain-rate at $Wi\approx 2/ 3$ when an EC regime is established. At the right of panel (b) this is observed only for $De=0.5$ and $De=1$. However, at $De=0.2$ the transition to an exponential thinning can barely be detected. Thus, there appears to be a critical value of the Deborah number in the range  $0.2 < De \leq 0.5$ beyond which an EC regime is established. Careful inspection of Figure \ref{fig:results_De} (b) also reveals that the presence of viscoelasticity, characterised by any finite $De$, leads to a slightly accelerated initial thinning in the IC regime in comparison to the Newtonian case (solid black line) for which $De=0$.

\begin{figure}[htb!]
\centering
  \includegraphics[height=6 cm]{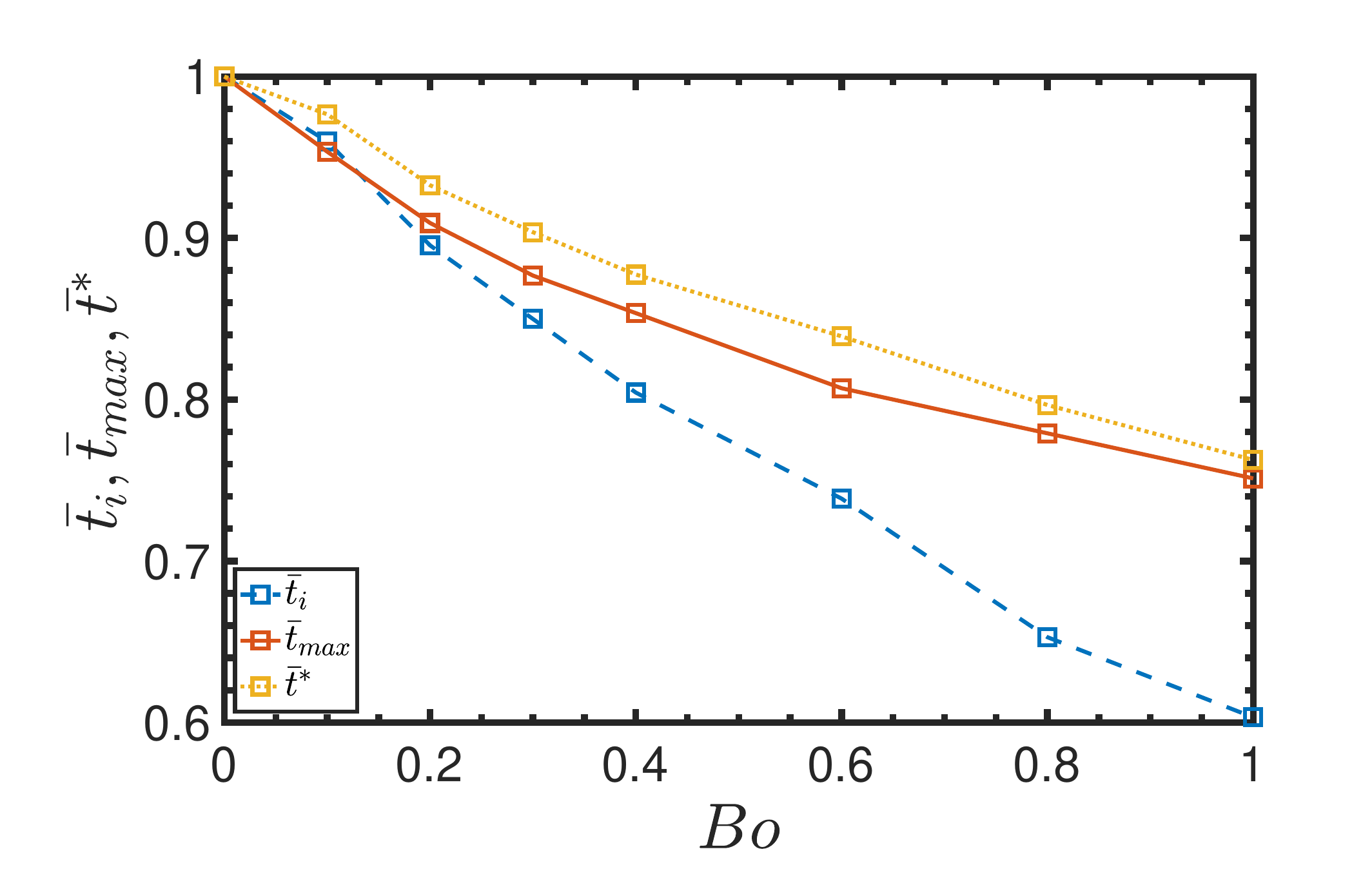}
  \caption{Variation of the scaled characteristic times $\Tilde{t}_i$, $\Tilde{t}_{max}$ and $\Tilde{t}^*$ with Bond number. Here, we consider the zero gravity case ($Bo=0$) as a reference to quantify the effect of gravitational forces (finite Bond numbers) on the onset of the inertio-capillary thinning ($\Tilde{t}_i$), the subsequent transition to the Elasto-Capillary (EC) regime ($\Tilde{t}_{max}$) until the final filament breakup ($\Tilde{t}^*$)}.
  \label{fig:results_Time_Bo}
\end{figure}
\begin{figure*}[ht!]
 \includegraphics[width=\textwidth]{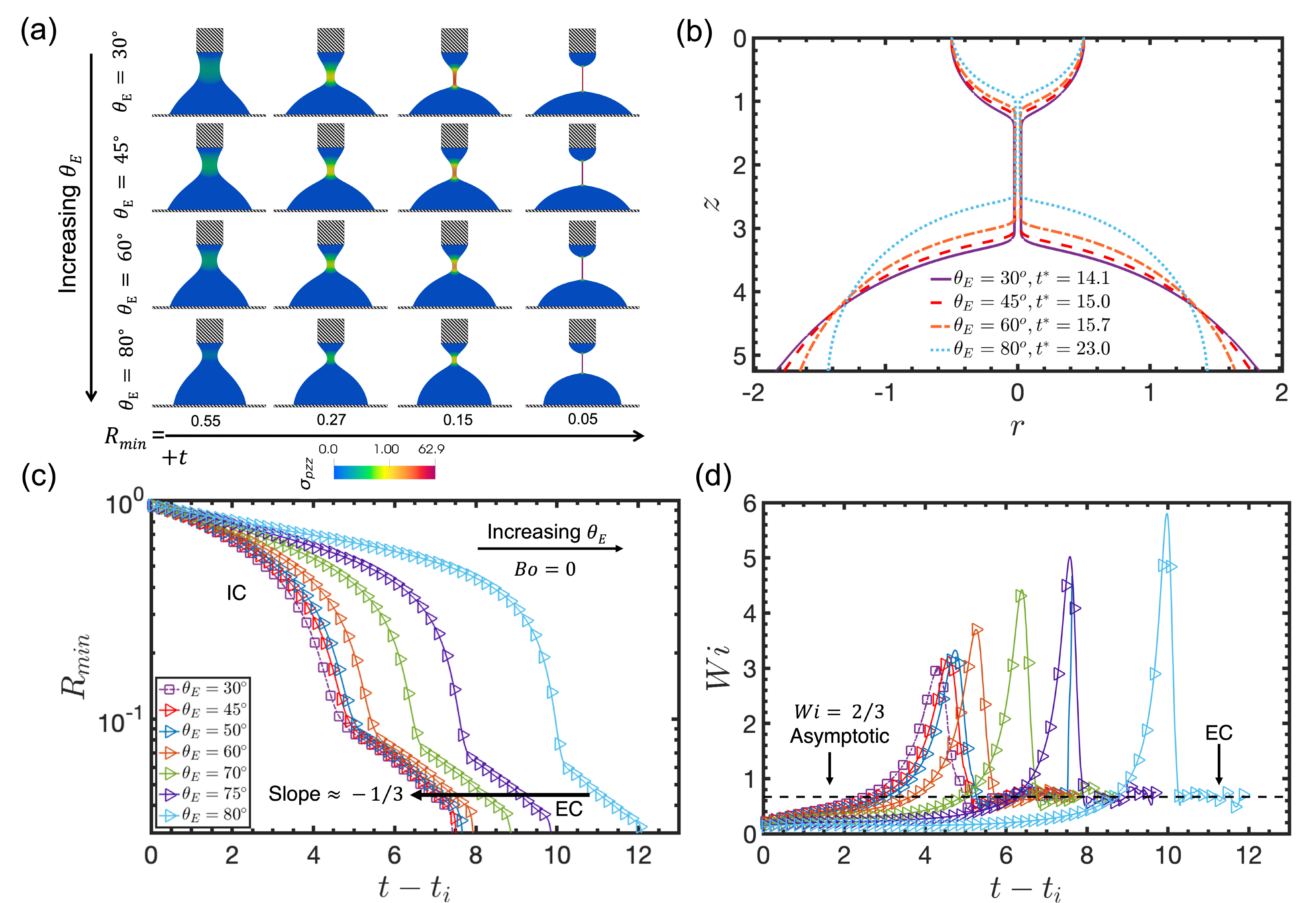}
  \caption{(a) Contour plots of the axial stress component at $R_{min} = 0.55$, 0.27, 0.15 and 0.05 for four contact angles $\theta_{E} = 30^\circ$, $45^\circ$, $60^\circ$ and $80^\circ$, covering different levels of substrate wettability. (b) The filament interface at times $t^*$ when $R_{min}=0.05$ for the same range of contact angles. (c) Evolution in the minimum filament radius $R_{min}(t-t_i)$ and (d) evolution in the local dimensionless strain rate $Wi(t)$ for substrate wettabilities in the range of $30^\circ \leq \theta_{E} \leq 80^\circ$.  The rest of the parameters remain unchanged from Table \ref{table_DoS}. For the corresponding movie please see the Electronic Supporting Information.}
  \label{fig:Theta_results}
\end{figure*}
Having identified the threshold $De$ above which effective measurements of the elastic properties of polymeric solutions are possible, we now examine the role of the chain finite extensibility parameter $L^2$. Figure \ref{fig:results_L2} shows that higher values of $L^2$ permit the development of larger axial polymeric stresses during the EC regime. However, there are no apparent differences in the interface evolution and the thread thickness, which is confirmed by inspection of Figure \ref{fig:results_L2}(b). The temporal evolution of the filament radius $R_{min}$ in Figure \ref{fig:results_L2}(c) shows that for small $L^2$ values ($L^2=400$ and 900), a much steeper (but still approximated exponential) decrease of the filament radius is observed. The profile of the corresponding local dimensionless strain-rate $Wi$ in Figure \ref{fig:results_L2}(d) confirms the faster rate-of-thinning of the filament radius for $L^2=400$ and $L^2=900$, which correspond to higher $Wi$ values ($Wi \geq 1$). For these relatively small values of the polymer finite extensibility, the stretching limit of the polymeric chains in the thinning thread is approached rapidly soon after the onset of the EC regime, which consequently results in a very short exponential thinning period. In contrast, the duration of the exponential filament thinning regime increases with $L^2$ and at sufficiently high extensibilities (here we regain the Oldroyd-B limit $Wi = 2/3$ for $L^2=10000$). These results demonstrate that a sufficiently large chain extensibility is needed for accurate rheological measurements in DoS rheometers. 


\begin{figure}[htb!]
\centering
  \includegraphics[height=6 cm]{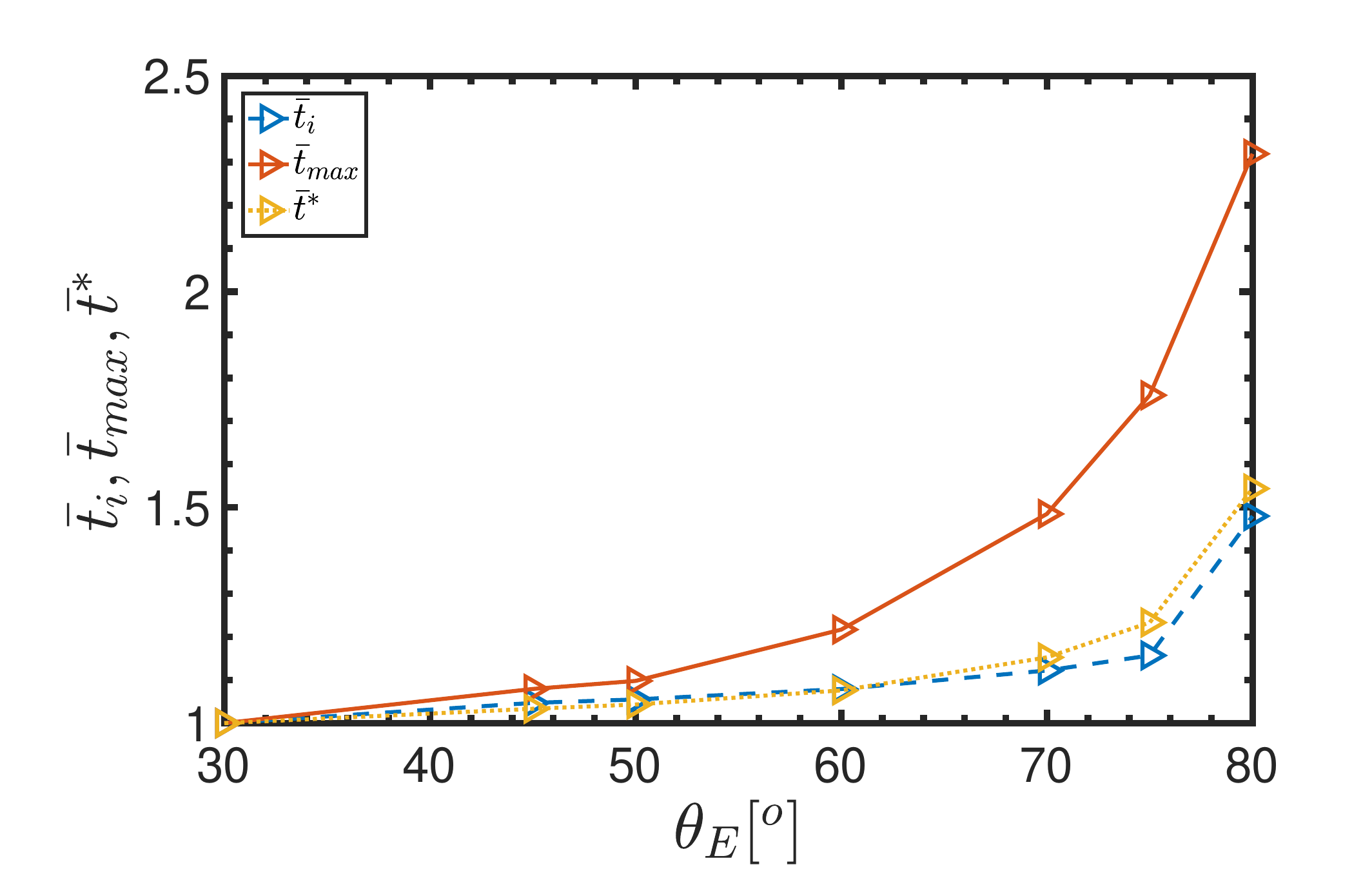}
  \caption{Variation of the scaled characteristic times $\Tilde{t}_i$, $\Tilde{t}_{max}$ and $\Tilde{t}^*$ with the static contact angle $\theta_E$. Here, we consider the case of the fastest substrate spreading ($\theta_E = 30 ^ \circ$) as a reference to quantify the effect of substrate wettability on the initial onset of the inertio-capillary thinning ($\Tilde{t}_i$), the subsequent transition to the Elasto-Capillary (EC) regime ($\Tilde{t}_{max}$) and the approach to the final filament breakup ($\Tilde{t}^*$).}
  \label{fig:results_Time_Theta}
\end{figure}

We now study the effect of gravity on the interfacial dynamics during dripping onto a substrate. Figure \ref{fig:Bo_results}(a) presents contour plots of the axial polymeric stress component for $Bo=0$, 0.1, 0.4, and 0.8, respectively, with a constant fluid contact angle $\theta_E=30^\circ$. Inspection of this figure reveals that the development of the elastic stress is only weakly dependent on gravity over the range of $Bo$ values that we examine. Furthermore, a quantitative comparison of the filament profiles as shown in Figure \ref{fig:Bo_results}(b) demonstrates that increasing $Bo$ leads to slightly longer filaments and an increase in the drop footprint across the solid substrate. In addition, Figures \ref{fig:Bo_results}(\textit{c}) and (\textit{d}) show that even though an increase in the gravitational body force leads to faster thinning in the IC regime as well as a more rapid transition to the onset of the characteristic exponential thinning regime, the dynamics in the EC regime are not influenced by variations in the Bond number. These results demonstrate that DoS rheological measurements of viscoelastic fluid properties, such as the relaxation time, remain unaffected by gravitational perturbations in the range $0 \leq Bo \leq 1$.

To better understand and quantify the influence of gravity on the thinning dynamics, it is worth examining the dependence of important features of the filament thinning profile, such as the characteristic times $t_i$ (which indicates when initial geometric effects become negligible), $t_{max}$ (which corresponds to the local maximum filament strain rate and the subsequent onset of the EC regime), and $t^*$ (which represents the thinning time, when the thread can be considered to be fully established and is about to undergo breakup, here when $R_{min}=0.05$). The dependence of $t_i$, $t_{max}$, and $t^*$ on $Bo$ is presented in Table \ref{table_t_Bo}.

In addition, in Figure \ref{fig:results_Time_Bo} we show the variation of these characteristic times with Bond number. Specifically, we plot the scaled quantities $\Tilde{t}_i = t_i (Bo) /t_i (Bo=0)$, $\Tilde{t}_{max} = t_{max} (Bo) / t_{max} (Bo=0)$, $\Tilde{t}^* = t^* (Bo) / t^* (Bo=0)$ where we use the zero Bond number case as a reference to determine the effect that gravitational forces exert on the initial phase ($t_i$) in DoS rheometry, the transition from the inertio-capillary to elasto-capillary dynamics ($t_{max}$) and finally to the time of the ``pinch-off" singularity ($t^*$). Figure \ref{fig:results_Time_Bo} confirms the initial observation that gravity accelerates the filament thinning, where the onset of the power-law and exponential filament thinning regimes at larger Bond numbers ($Bo \approx 1)$ are up to $40 \%$ and $20 \%$ faster, respectively.

\begin{table}[hbt!]
\centering
\caption{Dependence of the characteristic $t_i$, $t_{max}$ and $t^*$ timescales on the Bond number. These times correspond to the onset of the Inertio-Capillary (IC) regime, the onset of the elasto-capillary dynamics, and the time when the minimum filament radius decreases to $5 \%$ of its initial value ($R_{min}(t^*) \rightarrow 0.05$).}
\begin{tabular}{c c c c}
\hline
$Bo$ & $t_i$ & $t_{max}$ & $t^*$\\  [0.5ex]
\hline
0 & 7.46 & 11.75 & 14.11\\
0.1 & 7.16 & 11.27 & 13.78\\
0.2 & 6.68 & 10.59 & 13.16\\
0.3 & 6.34 & 10.10 & 12.75\\
0.4 & 6.01 & 9.68 & 12.38 \\
0.6 & 5.51 & 8.99 & 11.84\\
0.8 & 4.87 & 8.22 & 11.24\\
1 & 4.51 & 7.74 & 10.76\\
\end{tabular}
\label{table_t_Bo}
\end{table}


The solid surface wettability, characterised by the macroscopic contact angle $\theta_{E}$, influences the lateral spreading of the drop that is deposited on the substrate which, in turn, can play a role in driving the filament thinning dynamics. We study partially wetting solid substrates with wettabilities characterised by $\theta_{E}=30^\circ, ~45^\circ,~60^\circ,~{\rm and}~ 80^\circ$, which correspond to hydrophilic and partially hydrophobic substrates at the lower and higher ends of the contact angle range, respectively. Here, we have also set $Bo=0$ to isolate the effects of wettability from those associated with gravity. The contour plots in Figure \ref{fig:Theta_results}(a) show that increasing substrate wettability (by decreasing $\theta_{E}$) leads to an enhancement in the degree of spreading along the substrate, resulting in larger drop footprints and also longer connecting filaments. In contrast, more hydrophobic substrates lead to bead-like drops with smaller footprints and shorter connecting filaments, as can also be seen in Figure \ref{fig:Theta_results}(b). However, in contrast to the gravitational effects discussed above, the increase in $\theta_{E}$ is seen to retard significantly the onset of the EC regime for $\theta_{E} \geq 60^{\circ}$, as shown in Figure \ref{fig:Theta_results}(c) and (d), although once established, the exponential EC thinning regime retains the theoretically expected rate-of-thinning of $-1/(3 De)$ for all $\theta_{E}$ values studied.

\begin{table}[h!]
\centering
\caption{Dependence of the characteristic timescales $t_i$, $t_{max}$ and $t^*$ on the substrate wettability (represented by $\theta_{E}$).}
\begin{tabular}{c c c c}
\hline
$\theta_{E}[^\circ]$ & $t_i$ & $t_{max}$ & $t^*$\\  [0.5ex]
\hline
30 & 7.46 & 11.75 & 14.11\\
45 & 7.81 & 12.44 & 14.58\\
50 & 7.87 & 12.58 & 14.72\\
60 & 8.05 & 13.27 & 15.18\\
70 & 8.37 & 14.74 & 16.26\\
75 & 8.63 & 16.18 & 17.40\\
80 & 11.04 & 20.99 & 21.78\\
\end{tabular}
\label{table_t_theta}
\end{table}

Similarly to our analysis of the gravitational effects, here we also examine the impact of decreasing substrate wettability on the characteristic timescales $t_i$, $t_{max}$ and $t^*$. The results for these three timescales as a function of the contact angle $\theta_{E}$ are provided in Table \ref{table_t_theta}. In addition, Figure \ref{fig:results_Time_Theta} shows the evolution of the scaled characteristic times $\Tilde{t}_i = t_i (\theta_E) /t_i (\theta_E=30 ^ \circ)$, $\Tilde{t}_{max} = t_{max} (\theta_E)  / t_{max} (\theta_E=30 ^ \circ)$, $\Tilde{t}^* = t^* (\theta_E) / t^* (\theta_E=30 ^ \circ)$, selecting the lowest contact angle ($\theta_E = 30 ^ \circ$) as the reference case for quantifying the role of substrate wetting. Figure \ref{fig:results_Time_Theta} confirms the observation of Figure \ref{fig:Theta_results}: The use of more hydrophobic substrates causes significant retardation of the thinning dynamics, including the duration of the initial geometric rearrangement, the transition to the exponential EC region, and the final breakup of the viscoelastic thread. The trends presented in Figure \ref{fig:results_Time_Theta} show that increasing the contact angle $\theta_E$ results in an opposite trend from the increase in Bond number. There is a significant delay in the transition to the EC regime, of up to $23 \%$ for more hydrophobic substrates. 

These results demonstrate that the operating range in DoS rheometry can be increased by tuning the substrate wettability. Specifically, increasing the substrate wettability leads to a more rapid initial necking during the IC regime. The capillarity-driven drainage of fluid away from the thinning neck leads to more rapid growth of the elastic stresses in the thinning thread and an earlier onset of the EC regime (at slightly larger length scales). This is the configuration we seek to establish in DoS rheometry because it corresponds to a constant imposed strain rate and facilitates determination of the (unknown) relaxation time of a single fluid.

\section{Extensional Rheometry with DoS \label{sec:Fitting}}

 In this section, we explore conditions for optimising the operational range of DoS rheometry. We also present a simplified one-dimensional model of the thinning dynamics whose predictions are compared to those from the full numerical simulations. Finally, by comparing our results to the predictions of the FENE-P model in a capillary thinning flow we provide a methodology for improving the analysis of transient extensional rheometry data obtained from DoS experiments near the limit of finite time breakup.

\subsection{One-dimensional thinning model}
We begin with the development of a one-dimensional model for describing the thinning dynamics in a DoS flow configuration. This is inspired by the one-dimensional models developed by \citet{Tirtaatmadja2006} and \citet{Wagner2015}, but we seek to extend these descriptions to incorporate the perturbative roles of gravitational body forces and substrate wettability. We first consider the force balance for a slender viscoelastic thread expressed in dimensionless form \citep{Tirtaatmadja2006}:
\begin{equation}
  \dfrac{1}{R_{min}(t)} = C_1 \left(-\dot{R}_{min}(t)\right)^{2} + Oh\left[6 \beta  \dfrac{\dot{R}_{min}(t)}{R_{min}(t)} + \left( 1-\beta \right) \dfrac{1}{De} \Delta \sigma_p(t)\right],
  \label{eq:force-balance_scaled}
\end{equation}
where the over-dot designates a total derivative with respect to time. In Eq. (\ref{eq:force-balance_scaled}), the term on the left-hand side corresponds to the capillary pressure contribution driving the thinning. On the right-hand-side, 
the first term describes inertial acceleration in which $C_1$ is an adjustable pre-factor which captures the effects of gravitational body forces and substrate wettability on the initial inertio-capillary regime (as reported in Section \ref{sec:Results}). 
The second term on the right-hand-side of Eq. (\ref{eq:force-balance_scaled}) expresses the contribution of viscous forces, where $\beta = \eta_s / \eta_l$ is the relative contribution of the Newtonian viscous solvent, and the last term, given by
  \begin{equation}
    \Delta \sigma_p(t) = \left[ A_{zz} - A_{rr}\right]/\left({1 - {tr\left(\mathbf{A} \right)}/{L^2}} \right),  
  \end{equation}
corresponds to the dimensionless polymeric normal stress difference that develops in the thinning filament.  

If the contribution of  the viscous stresses is negligible ($Oh \ll 1$) compared to the capillary and elasticity contributions, then  Eq. (\ref{eq:force-balance_scaled}) can be re-written:
%
\begin{equation}
\dfrac{1}{R_{min}(t)} \sim  C_1 \left(-\dot{R}_{min}(t)\right)^{2}  + \left( 1-\beta \right) \dfrac{Oh}{De}  \dfrac{\left( A_{zz} - A_{rr}\right)}{\left[1 - \frac{tr\left(\mathbf{A} \right)}{L^2}\right]}. 
  \label{eq:force-balance_scaled2}
\end{equation}
As the filament necks down, we have seen from our simulations (cf. Figure \ref{fig:results_De}(a)) that the axial stress component eventually becomes dominant, so that $A_{zz} \gg A_{rr}$. Furthermore, in the limit of large polymer chain extensibility, before the effects of finite extensibility become important, we can take $1/\left(1-tr\left(\mathbf{A}\right)/L^2\right) \approx 1$. Thus, provided $1 \leq A_{zz} \ll L^2$, Eq. (\ref{eq:scaled-conformation}) combined with Eq. (\ref{eq:Wi}) reduces to a single evolution equation for $A_{zz}(t)$ and $R_{min}(t)$: 
%
\begin{equation}
    \dot{A}_{zz}\approx -\left(4\frac{\dot{R}_{min}(t)}{R_{min}(t)}+\frac{1}{De}\right)A_{zz}.
\end{equation}
Separation of variables and integration of this equation yields an expression for $A_{zz}$ in terms of $R_{min}(t)$ and $De$ \citep{Entov1997}:
\begin{equation}
  A_{zz}(t) = \dfrac{A_{zz}^0}{R_{min} ^4(t)} \exp{\left( -\frac{t}{De}\right)},
  \label{eq:Azz}
\end{equation}
where $A_{zz}^0$ is an integration constant that can be set to unity corresponding to an initially unstretched polymer chain. 
Substituting Eq. (\ref{eq:Azz}) into Eq. (\ref{eq:force-balance_scaled2}), leads to the following nonlinear evolution equation for $R_{min}(t)$:
\begin{equation}
\dfrac{1}{R_{min}(t)} \approx C_1 \left(-\dot{R}_{min}(t)\right)^{2} + \left( 1-\beta \right) \dfrac{Oh}{De}  \dfrac{1}{R_{min}^4 (t)} \exp{\left( -\frac{t}{De}\right)}.
  \label{eq:force-balance_1D}
\end{equation}
%

Equation (\ref{eq:force-balance_1D}) incorporates the competing effects of fluid inertia, capillarity, and nonlinear fluid elasticity on the dynamics of filament thinning. At very short times, before elastic stresses grow to enter the dominant balance, balancing just the capillary and inertial terms in the IC regime, we have $1/R_{min}(t) \approx C_1 (-\dot{R}_{min}(t))^2$. Integration of this equation from $R_{min}=1$ at $t=0$ results in the expression:
\begin{equation}
    R_{min}(t) = \alpha (t_{max}-t)^{2/3}.
    \label{eq:IC}
\end{equation}

The radius therefore initially decreases as a power law in time. Here, $\alpha$ is a dimensionless pre-factor which determines the rate-of-thinning during the inertio-capillary balance \citep{Day1998, Eggers2015, Dinic2019, MartinezNarvaez2021} and is related to the pre-factor $C_1$ in the original force balance by the expression $C_1=9/ (4 \alpha^3)$. In the absence of fluid elasticity $t_{max}$ is the time to breakup; however substituting Eq. (\ref{eq:IC}) into Eq.(\ref{eq:Azz} it is clear that the polymer stretch grows extremely rapidly during this inertio-capillary thinning phase until elastic stresses become large enough to enter the dominant balance in Eq. (\ref{eq:force-balance_scaled2}). At this time (denoted $t_{max}$), the thinning rate drops dramatically, inertial effects become progressively less important and the transition to the EC regime begins.
In the EC regime, the dominant balance between capillarity and elasticity in Eq.(\ref{eq:force-balance_1D}) gives:
\begin{equation}
    \dfrac{1}{R_{min}(t)} \approx (1-\beta)\left(\dfrac{Oh}{De}~\dfrac{1}{R_{min}^4 (t)} \right)\exp(-t/De),
\end{equation}
Rearranging we obtain the following expression for the evolution of the filament radius in the EC regime:
\begin{equation}
    R_{min}(t) \approx \left[(1-\beta)\dfrac{Oh}{De}\right]^{1/3}\exp(-t/3De),
\end{equation}
which is consistent with the characteristic exponential rate-of-thinning in the viscoelastic filament confirmed notably by \citet{Entov1997}, \citet{Amarouchene2001}, \citet{Deblais2020}. At very long times, as the filament thins to very small scales, $A_{zz}(t)$ approaches the finite extensibility limit $L^2$ and the thread enters the TVEC regime \citep{MartinezNarvaez2021}. In this regime, the radius once again decreases linearly in time and the strain rate grows rapidly. The dynamics of this regime have been considered by \citet{Wagner2015} but are very challenging to resolve numerically and are beyond the scope of the simulations in the present paper. 

This simple one-dimensional model captures the overall dynamics of the filament thinning process but is not complete enough to describe the additional driving forces provided by the perturbative effects of gravity or substrate wettability. However, the contribution of these effects to the inertio-capillary thinning process (observed in Figures \ref{fig:Bo_results}  and \ref{fig:Theta_results}) can be captured through the changes in the pre-factor $\alpha$ defined in Eq. (\ref{eq:IC})
\begin{figure}[htb!]
  \includegraphics[height=6.75cm]{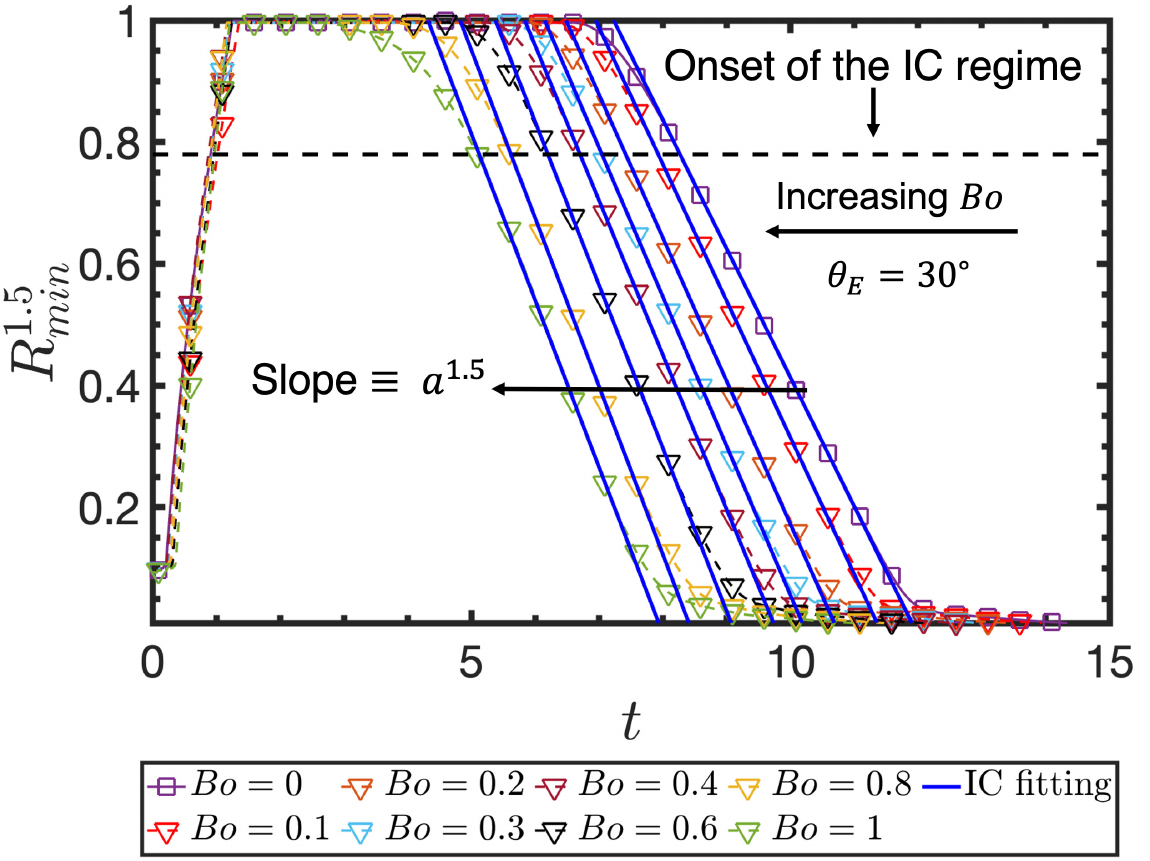}
  \caption{Temporal variation of $R^{1.5}_{min}$ for $0 \leq Bo \leq 1$ and the corresponding linear fits (blue solid lines) allow determination of the pre-factor $\alpha$, according to the power-law associated with the inertio-capillary dynamics given by Eq. (\ref{eq:IC}). The initial onset of the IC thinning regime can be identified to be at $R^{1.5}_{min} \approx 0 .8$ (or equivalently $R_{min} \approx 0.85$) (dashed line).}
  \label{fig:IC_alpha}
  \end{figure}

To illustrate this, following the graphical approach suggested by \citet{Day1998} we use a re-arranged form of Eq. (\ref{eq:IC}), $R_{min}^{3/2}(t) = -\alpha^{3/2}t + \alpha^{3/2}t_{max}$, and we plot in Figure \ref{fig:IC_alpha} the temporal evolution of $R_{min}^{3/2}$ on a semi-log scale; this allows us to rapidly determine the value of the constant $\alpha$ for each simulation and how it varies with the Bond number. It is clear that for all $Bo$ examined, there is a distinct range $0.1 \leq R_{min} ^{3 /2} \leq 0.8$, where $R_{min}^{3/2}$ decreases linearly in time, consistent with the IC regime. We indicate the onset of the IC regime at a specified time, named $t_{IC}$ which corresponds to $R_{min} \approx 0.85$ (or equivalently $R_{min} ^{3/2} \approx 0.8$), which is a value that can be easily monitored in the experiments. Fitting of the linear portions of the data in Figure \ref{fig:IC_alpha} accounting for the re-arranged form of Eq. \ref{eq:IC}, yields $0.355 \leq \alpha \leq 0.425$ for $0\leq Bo \leq 1$.

We follow the same procedure for quantifying the effect of the substrate wettability on the inertio-capillary thinning, which yields  $0.355 \leq \alpha \leq 0.212$ for $30^{\circ} \leq \theta_E \leq 80^{\circ}$. We provide the values of the pre-factor $\alpha$ as a function of the Bond number and the contact angle $\theta_E$ in Tables \ref{table_alpha_Bo} and \ref{table_alpha_theta}, respectively. 

\begin{figure}[htb!]
  \includegraphics[height=11.0cm]{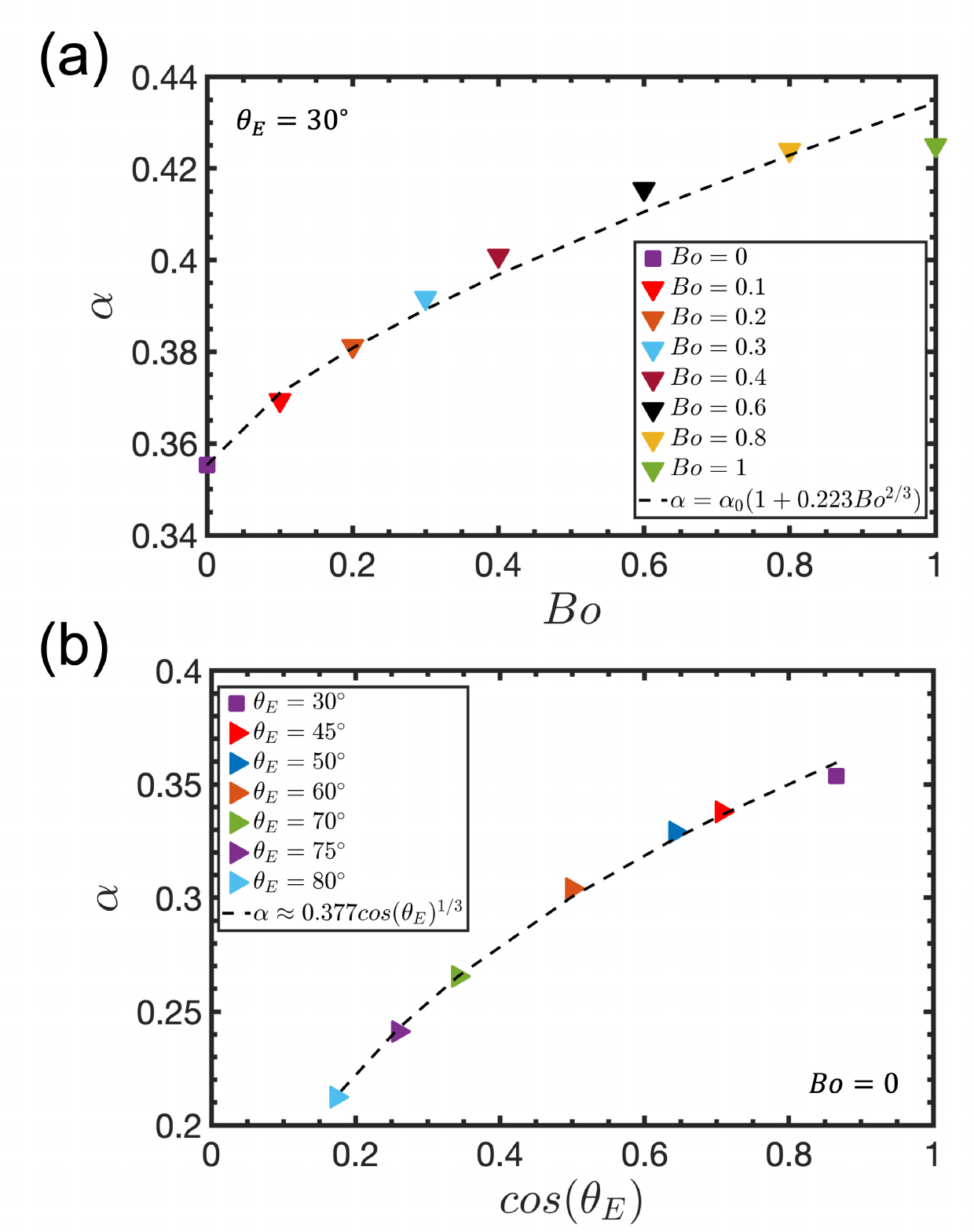}
  \caption{Variation of the pre-factor $\alpha$ with (a) the Bond number and (b) the substrate wettability characterised by $\cos(\theta_{E})$, is characterised by a power law fit of index 2/3 and 1/3, respectively.}
  \label{fig:alpha}
  \end{figure}

In addition, we also plot in Figure \ref{fig:alpha}(a) and (b) the variation of this numerical pre-factor $\alpha$ as a function of the Bond number and the cosine of the contact angle ($cos\left(\theta_E\right)$), which provides a measure of the force driving the lateral spreading of the fluid across the solid substrate. The pre-factor $\alpha$ in Eq. (\ref{eq:IC}) follows a power-law evolution with $Bo$, which can be approximated by $\alpha(Bo) \approx \alpha_0 \left( 1 + 0.223 Bo^{2/3}\right)$, where $\alpha_0$ corresponds to the value of $\alpha$ for $Bo=0$. The inertio-capillary scaling of the initial thinning regime is thus only weakly modified by gravity (for $Bo \leq 1$). Inspection of Figure \ref{fig:alpha}(b) shows that as the wettability of the substrate is reduced the rate of spreading is also reduced. The dependence of $\alpha$ on the contact angle can be closely approximated by $\alpha(\theta_{E}) \approx 0.377 \cos(\theta_{E})^{1/3}$.

\begin{table}
\small
\caption{Variation of the pre-factor $\alpha$ for the Inertio-Capillary (IC) regime as a function of the Bond number.}
\begin{tabular}{c c c c c c c c c}
\hline
$Bo$ & 0 & 0.1 & 0.2 & 0.3 & 0.4 & 0.6 & 0.8 & 1 \\
\hline
$\alpha$ & 0.355 & 0.369 & 0.381 & 0.392 & 0.401 & 0.412 & 0.424 & 0.425 \\
\hline
\end{tabular}
\label{table_alpha_Bo}
\end{table}

\begin{table}
\small
\caption{Dependence of the pre-factor $\alpha$ for the Inertio-Capillary (IC) regime on the substrate wettability characterised by the contact angle $\theta_{E}$.}
\begin{tabular}{c c c c c c c c}
\hline
$\theta_{E}$ & $30^\circ$ & $45^\circ$ & $50^\circ$ & $60^\circ$ &
$70^\circ$ & $75^\circ$ & $80^\circ$ \\
\hline
$\alpha$ & 0.355 & 0.338 & 0.329 & 0.304 & 0.266 & 0.241 & 0.212 \\
\hline
\end{tabular}
\label{table_alpha_theta}
\end{table}\

\begin{figure}[htb!]
  \includegraphics[height=12.5cm]{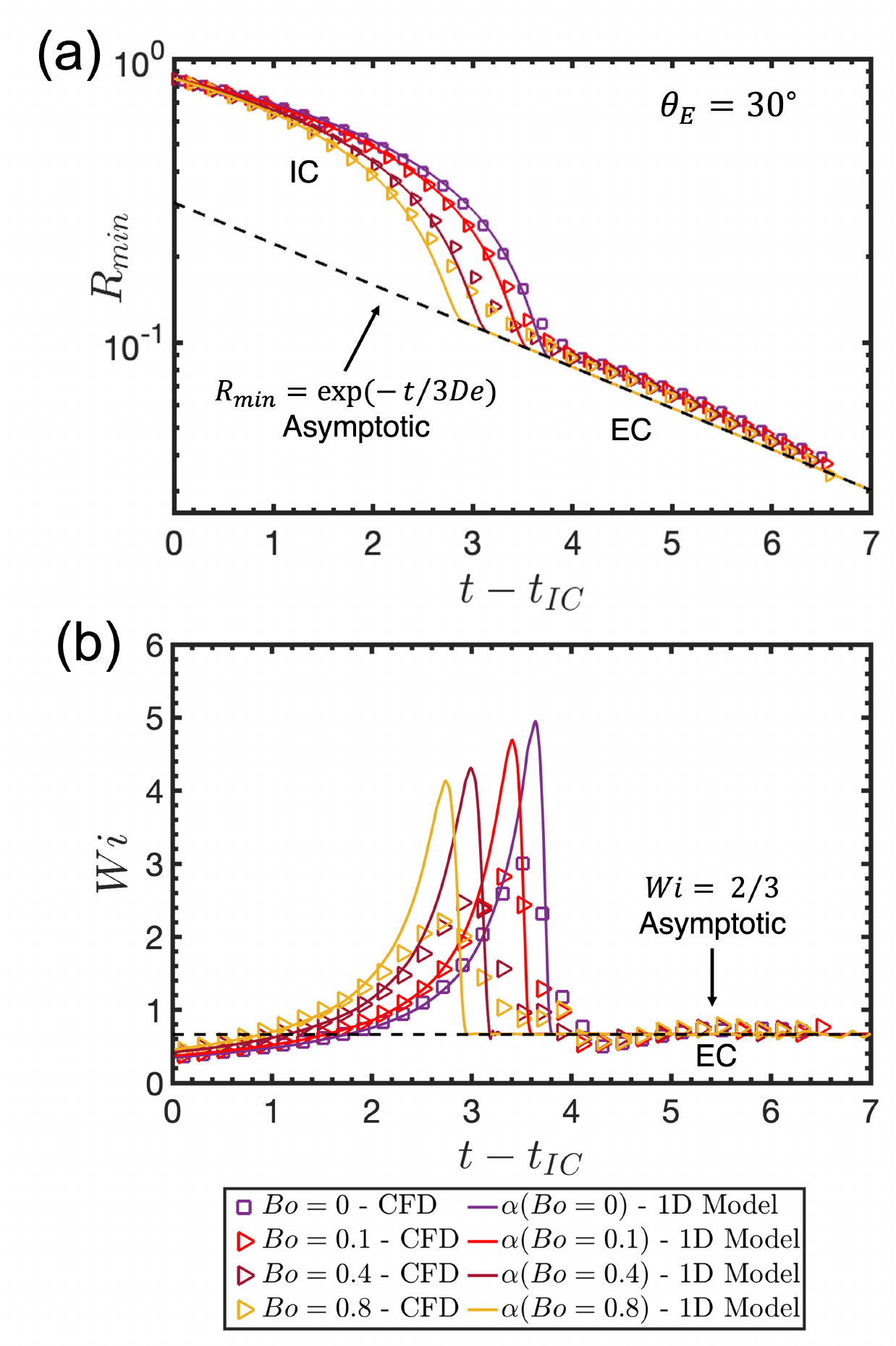}
  \caption{One-dimensional model predictions (solid lines) and numerical simulation results (symbols) for the temporal evolution of (a) the viscoelastic filament radius $R_{min}$ and (b) the dimensionless strain rate $Wi$ in the neck. Here $Bo$ varies as in Figure \ref{fig:Bo_results} and the rest of the parameters remain unchanged. The exponential filament thinning rate in the Elasto-Capillary (EC) regime is also shown in (a) and the corresponding constant $Wi = 2/3$ plateau (dashed lines) is shown (b).}
  \label{fig:1D_Rmin}
  \end{figure}

Having determined the variation of $\alpha$ with the Bond number and the macroscopic contact angle, we compare the predictions of the one-dimensional model in Eq. (\ref{eq:force-balance_1D}) with the results of the numerical simulations, for $0\leq Bo \leq 1$ at a fixed contact angle $\theta_E = 30^{\circ}$. We show in Figure \ref{fig:1D_Rmin}(a) and (b) the results of the comparison between the full numerical simulations and the solution of Eq. (\ref{eq:force-balance_1D}) (using the Matlab ODE23t package), for both the temporal evolution of the filament radius $R_{min}(t)$ and the corresponding evolution of the dimensionless strain rate $Wi(t)$, respectively. Here, the time $t$ has been shifted by an initial offset $t_{IC}$ which corresponds to the time when $R_{min}=0.85$ after the initial reconfiguration process (during which the prolate drop first touches the substrate and rapidly spreads laterally to form a necked droplet). Figure \ref{fig:IC_alpha} clearly shows that below a radius $R_{min} \approx 0.85$ (or $R_{min} ^{3/2} \approx 0.8)$ the initial phase of filament thinning is well described by the inertio-capillary scaling. Inspection of Figure \ref{fig:1D_Rmin}(a) and (b) reveals that the simplified one-dimensional model accurately captures both the IC and the EC thinning regimes. However, the transition between the two regimes is less abrupt in the full simulations as the shape of the filament rearranges and transient two-dimensional effects cannot be neglected. Similar observations have been made regarding the effect of the substrate wettability \citep{zinelis_thesis}.

\subsection{Improvement of thinning predictions with the FENE-P analytical solution}
\begin{table*}[htb!]
\small
\caption{Determination of the three fitting parameters in Eq. (\ref{eq:fenep}): Deborah number (De), elasto-capillary number (Ec), and polymeric finite extensibility ($L^2$); Here, $(De_{input}, Ec_{input}, L^2_{input})$ are the known (ground truth) values that are used as inputs to the numerical simulations;  $De^{[Oldroyd-B]}_{fit}$ and $De^{[AM]}_{fit}$ are the values of the dimensionless relaxation time obtained when we use the asymptotic Oldroyd-B result: $R_{min}(t) \sim \exp(-t/(3De)$ (labelled ``Oldroyd-B Fitting") and the ``Anna-McKinley" model provided in Eq. (\ref{eq:AM}) (labelled ``Anna-McKinley Fitting") to determine the dimensionless relaxation time of the dilute polymer solution from the data in the EC regime (here, we consider the data generated by the numerical simulations). $(De^{[FENE]}_{fit}, Ec^{[FENE]}_{fit}, L^2 {}^{[FENE]}_{fit})$ are the parameters obtained when Eq. (\ref{eq:fenep}) is fitted to the available data for the filament radius (labelled ``FENE-P Fitting''). The errors $\epsilon^{[Oldroyd-B]}$, $\epsilon^{[AM]}$ and $\epsilon^{[FENE]}$ are the discrepancies in the prediction of the dimensionless polymeric relaxation time $De_{fit}$ when either the Oldroyd-B result or Eq. (\ref{eq:AM}) or Eq. (\ref{eq:fenep}), respectively are used.}
\begin{tabular}{c c c |c|c|c c c|c c c}
\multicolumn{3}{c}{Input} & \multicolumn{1}{c}{Oldroyd-B Fitting} & \multicolumn{1}{c}{Anna-McKinley Fitting} &\multicolumn{3}{c}{FENE-P Fitting} & \multicolumn{3}{c}{Error for $De_{fit}$} \\
 \hline
$De_{input}$ & $Ec_{input}$ & $L^2_{input}$ & $De^{[Oldroyd-B]}_{fit}$ & $De^{[AM]}_{fit}$ & $De^{[FENE]}_{fit}$ & $Ec^{[FENE]}_{fit}$ & $L^2 {}^{[FENE]}_{fit}$ & $\epsilon^{[Oldroyd-B]}$ & $\epsilon^{[AM]}$ & $\epsilon^{[FENE]}$ \\  [0.5ex]
\hline
1 & 0.03 & 400 & 0.60 & $0.56 \pm 0.006$ & $0.70 \pm 0.006$ & 0.003 & $10^4$ & $-40\%$ & $-44\%$ & $-30\%$ \\
1 & 0.03 & 900 & 0.66 &  $0.80 \pm 0.001$ & $0.85 \pm 0.010$ & 0.430 &  $10^2$ & $-34\%$ & $-20\%$ & $-15\%$\\
1 & 0.03 & 1600 & 0.94  & $0.90 \pm 0.001$ & $0.91 \pm 0.006$ & 0.680 & $10^2$ & $-6\%$ & $-10\%$ & $-9\%$ \\
1 & 0.03 & 2500  & 1.00  & $0.96 \pm 0.007$ & $0.98 \pm 0.020$ &  0.740 & $10^2$ & 0\% & $-4\%$ & $-2\%$ \\
\end{tabular}
\label{table_FENE}
\end{table*}

We now revisit the role of the polymer finite extensibility $L^2$ on the dynamics in the EC thinning regime that is shown in Figure \ref{fig:results_L2}. Our simulations revealed that low values of the polymer extensibility ($L^2 <1600$) resulted in more rapid thinning rates in the EC regime and very short-lived periods of exponential thinning. Similar observations are obtained in experimental measurements by \citet{gaillard2023}. Directly fitting a simple exponential to such data leads to significant bias in DoS rheometry measurements of polymer solutions, such as an underestimation of the polymer relaxation time. The resulting error can be as large as approximately $40 \%$ for $L^2=400$ if the analytical expression for exponential filament thinning predicted by the Oldroyd-B model, i.e., $R_{min} \sim \exp(-t/(3De)$ \citep{Entov1997, Clasen2006, Rajesh2022}, is used to extract the thinning rate or the apparent relaxation time. To address this bias, \citet{Lauser2021} and \citet{Jimenez2020} use a four-parameter empirical model originally proposed by \citet{Anna2001} to more systematically regress experimental measurements of filament thinning close to breakup. In this model, the time-evolving neck radius is written in the form of a four-parameter fit:
\begin{equation}
    R_{min}(t-t_1)  = A \exp{(-B(t-t_1))} -C(t-t_1) + D,
    \label{eq:AM}
\end{equation}
where the dimensionless time $t$ is shifted by the dimensionless value $t_1$ which corresponds to the time at which the EC regime begins. The parameter $B$ is the thinning rate (expected to be $B = 1/ 3 De$) and the transition to a linear dependence in the final TVEC region close to the breakup is captured by the final terms ($C, D$)). \citet{Lauser2021} and \citet{Jimenez2020} show that a close agreement can be obtained between this functional form and experimental data. Here, we explore how well this formulation can also characterise the dynamics predicted by our numerical simulations (where the ``ground truth" is already known), as well as an alternative form motivated by the (known) analytic form of the one-dimensional filament thinning process for the FENE-P model.

\citet{Wagner2015} derived an implicit analytical expression describing the filament thinning and breakup of FENE-P fluids. Their expression accounts for the thinning dynamics in the EC and the terminal (TVEC) regimes where finite extensibility effects become important. In our notation this analytic solution can be written in the form:
\begin{multline}
  t-t_1 = -De \frac{\left(L^2 + 2\right)}{\left(L^2 + 3\right)^2} \Biggl\{ \frac{1}{1 + EcR_1\left(L^2 + 3\right)} - \frac{1}{1 + \xi EcR_1  \left(L^2 + 3\right)} \\
  + 3\ln\left(\frac{1 + \xi EcR_1\left(L^2 + 3\right)}{1 + EcR_1 \left(L^2 + 3\right)}\right) 
  + 4 Ec R_1  \frac{\left(L^2 + 3\right)}{\left(L^2 + 2\right)}\left(\xi - 1\right) \Biggr\}.
  \label{eq:fenep}
\end{multline}
 Here the time-evolving dimensionless filament radius $R_{min}(t)$ is re-scaled with the dimensionless value $R_1$ which represents the minimum filament radius at the onset of the  EC regime. Thus at time $t=t_1$ the re-scaled radius is $R_{min} / R_1 = 1$. This scaling leads to a new dimensionless filament radius $\xi = R_{min} /R_1$ and the dimensionless time ($t-t_1$) characterises the entire evolution of the EC thinning regime. In addition, $Ec=G R_0 / \gamma$ is the elasto-capillary number which provides the appropriate dimensionless measure of the fluid elasticity (i.e. the elastic modulus $G$ of the polymer solution) scaled with the capillary pressure $\gamma / R_0$. Equation (\ref{eq:fenep}) includes three independent fitting parameters: the relaxation time $\tau$ or (equivalently the Deborah number $De$ in dimensionless form), the elasto-capillary number $Ec$ and the finite extensibility $L^2$ of the dissolved polymer. In a DoS rheometry experiment with an unknown sample, we ideally seek to simultaneously determine all three fitting parameters. Because we know in advance the ground truth values of $De$, $Ec$ and $L^2$ (which act here as an input to our numerical simulations), we can explore the suitability of using Eq. (\ref{eq:fenep}) to accurately retrieve these three material parameters from measurements of the filament midpoint radius $R_{min}(t)$.

\begin{figure}[htb!]
  \includegraphics[height=5.8cm]{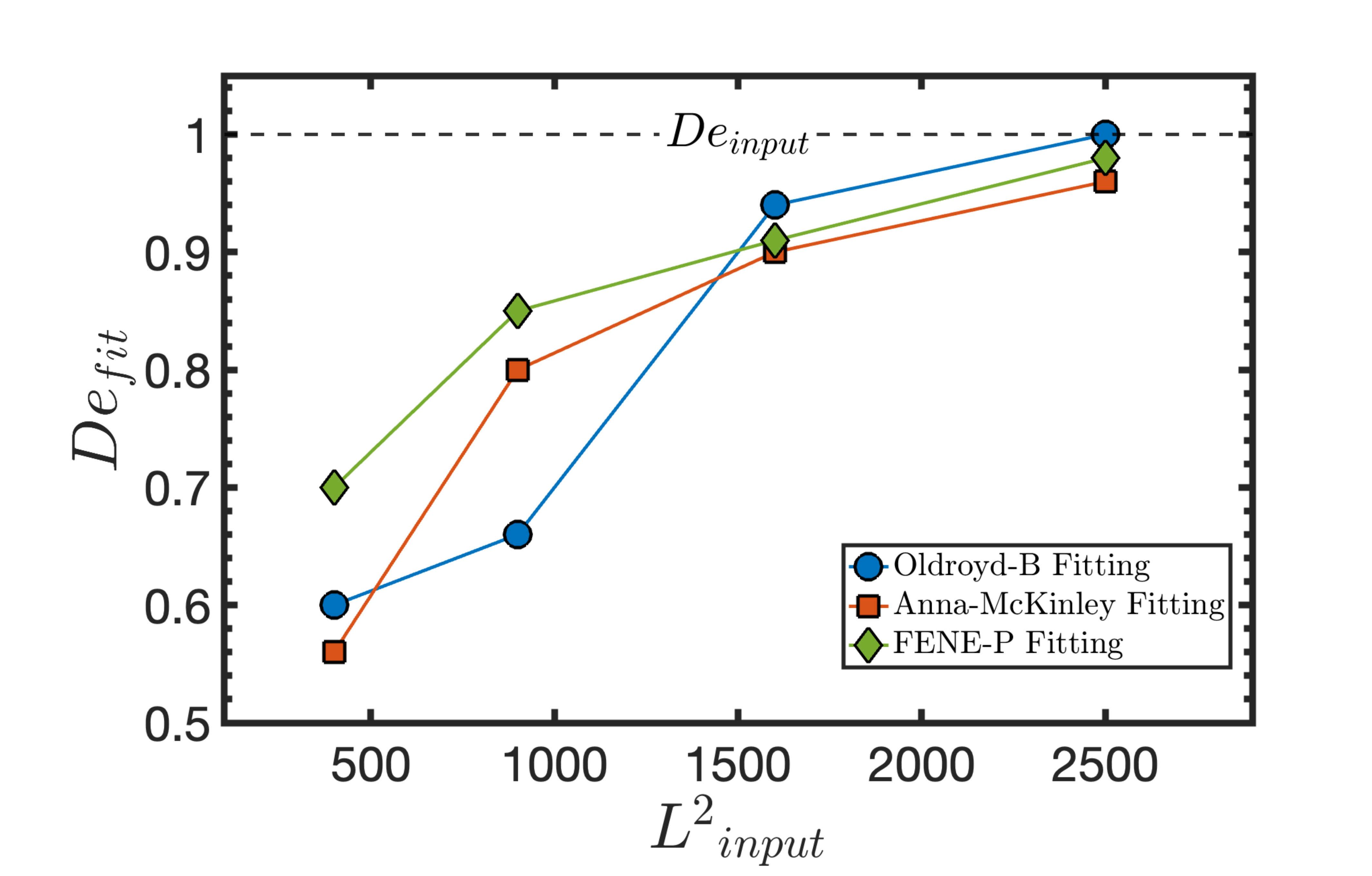}
  \caption{Estimated $De_{fit}$ values from Table \ref{table_FENE} for the range $L^2_{input} = \{400, 900, 1600, 2500 \}$ using the ``Oldroyd-B'' (blue circles), the ``Anna-McKinley'' (red squares) and the ``FENE-P'' (green diamonds) fitting approaches. The dashed black line shows the ground-truth value of $De_{input}=1$.}
  \label{fig:De_FIT}
  \end{figure}

 We provide in Table \ref{table_FENE} the results of the fitting process for each $(De_{input}, Ec_{input}, L^2_{input})$  to explore how accurately we can determine the (dimensionless) relaxation time by fitting the neck evolution during the thinning process. We consider first the well-known asymptotic Oldroyd-B result: $R_{min} \sim \exp{(-t/3De)}$ \citep{Entov1997, Clasen2006, Rajesh2022}, labelled ``Oldroyd-B Fitting" here. Subsequently, we fit the same data using the semi-empirical ``Anna-McKinley" model described by Eq. (\ref{eq:AM}). These results are labelled by the superscript $[AM]$. Finally, we also employ the FENE-P analytical solution provided in Eq. (\ref{eq:fenep}) to fit the 
 different computational datasets. This is labelled ``FENE-P Fitting''. Table \ref{table_FENE} highlights the high bias error obtained for low polymer extensibilities ($L^2_{input}= 400$ and 900) when the conventional ``Oldroyd-B Fitting" is performed. On the other hand, it is evident that for $L^2_{input} = 400$  the use of Eq. (\ref{eq:fenep}) reduces considerably the estimated error for $De_{fit}$ by approximately $10\%$. However, the use of Eq. \ref{eq:AM}) results in a slightly worse estimated error in the value determined for $De_{fit}$ compared to the ``Oldroyd-B Fitting". For a moderate value of extensibility ($L^2_{input}= 900$) the ``Anna-McKinley Fitting" and ``FENE-P Fitting" both result in reductions in the error incurred in determining the fluid relaxation time by $14 \%$ and $19\%$, respectively. Surprisingly, however, these two methods of analysing Dripping-Onto-Substrate thinning data are both found to lead to marginally worse predictions of $De_{fit}$ for $L^2_{input}= 1600$ and 2500. We graphically summarise in Fig. \ref{fig:De_FIT} the estimated values of $De_{fit}$ using the ``Oldroyd-B", ``Anna-McKinley" and ``FENE-P" methods of analysis. It is evident that using the ``FENE-P" solution provides more accurate results than the ``Anna-McKinley" expression in all cases. In particular, it is seen that for low extensibility fluids ($L^2_{input}\leq900$) using Eq. (\ref{eq:fenep}) provides the most accurate determination of the characteristic fluid relaxation time. 
 
\begin{figure}[htb!]
  \includegraphics[height=5.8cm]{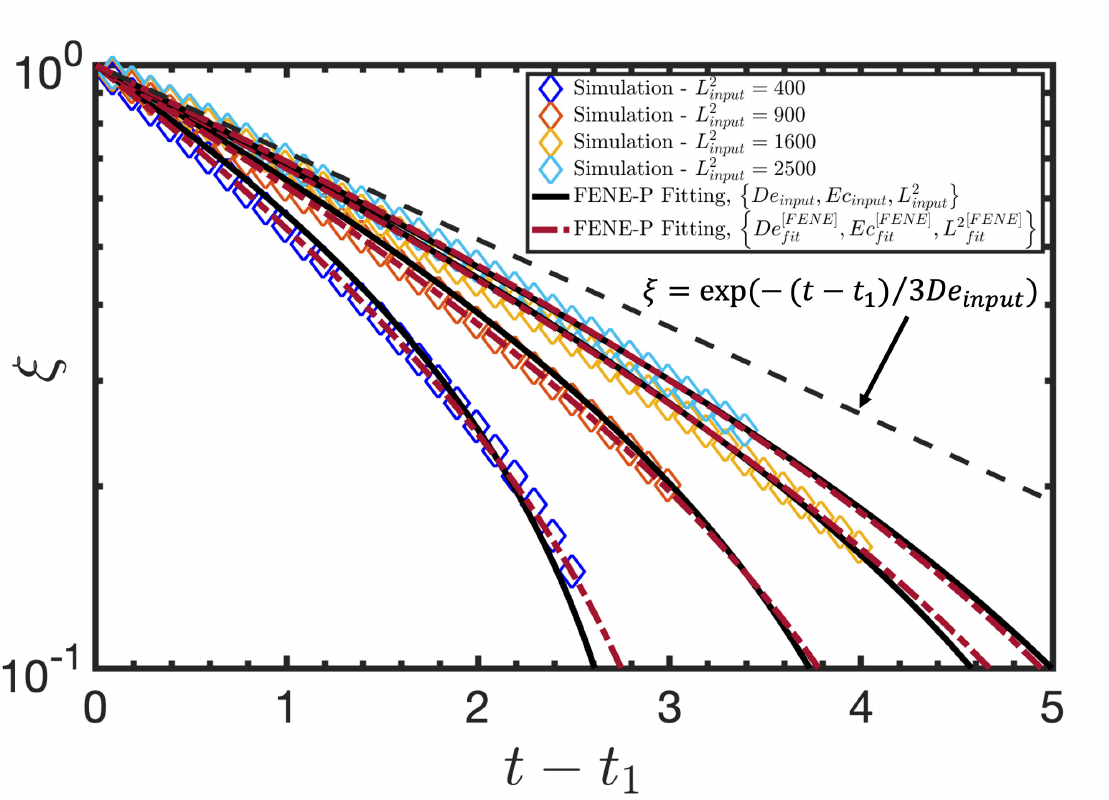}
  \caption{Simulation data denoted by diamond symbols for different values of polymer finite extensibility $L^2_{input} = \{400, 900, 1600, 2500 \}$ and fixed $De_{input}=1$, $Ec_{input}=0.03$. The solution of Eq. (\ref{eq:fenep}) for the same values $(De_{input}, Ec_{input})$ and range $L^2_{input} = \{400, 900, 1600, 2500 \}$ as in the numerical simulations are shown by solid black lines. Dashed red lines correspond to the predictions of Eq. (\ref{eq:fenep}) where the results of the ``Improved fitting" columns in Table \ref{table_FENE} for $(De^{[FENE]}_{fit}, Ec^{[FENE]}_{fit}, L^2 {}^{[FENE]}_{fit})$ are considered. The prediction of the exponential Oldroyd-B decay during the Elasto-Capillary (EC) regime \citep{Entov1997, Clasen2006, Rajesh2022} is also indicated by the dashed black line. }
  \label{fig:FENE_L2}
  \end{figure}

We now focus in more detail on the ``FENE-P Fitting" approach. Careful inspection of Table \ref{table_FENE} shows that using this fitting method leads to values of $Ec^{[FENE]}_{fit}$ and $L^2 {}^{[FENE]}_{fit}$ with pronounced deviations from the corresponding input values. We investigate this surprising result further by first plotting in Figure \ref{fig:FENE_L2} the evolution of the dimensionless filament radius $\xi(t-t_1) = R_{min}(t-t_1) / R_1$ with time $t-t_1$ to focus on the thinning during the elasto-capillary regime. Specifically, we plot (i) the results of the numerical simulations (symbols) for the four different values $L^2_{input} = \{400, 900, 1600, 2500\}$ at fixed $De_{input} = 1$ and $Ec_{input}=0.03$, (ii) the predictions of Eq. (\ref{eq:fenep}) for ${De_{input}, Ec_{input}, L^2_{input}}$ (solid black lines), and (iii) the results of Eq. (\ref{eq:fenep}) for $(De^{[FENE]}_{fit}, Ec^{[FENE]}_{fit}, L^2 {}^{[FENE]}_{fit})$ (dashed red lines). Inspection of Figure \ref{fig:FENE_L2} highlights the significant deviation from the simple exponential Oldroyd-B asymptotic result ($\xi \approx \exp(-(t-t_1)/3De)$) due to the effects of polymer finite extensibility. This deviation is most pronounced for $L^2 _{input}=400$, for which the exponential region is very short and restricted to times $t-t_1 \lesssim 1$. As a consequence of this truncated exponential thinning, robust estimation of the characteristic relaxation time for a dilute polymer solution with molecules of limited extensibility is very challenging. Moreover, we also observe that both the solid black and red dashed lines overlap substantially and both accurately capture the results of the numerical simulations (symbols), with significant deviations only being observed for the lowest extensibility case ($L^2_{input}=400$). 

\begin{figure}[htb!]
  \includegraphics[height=5.2cm]{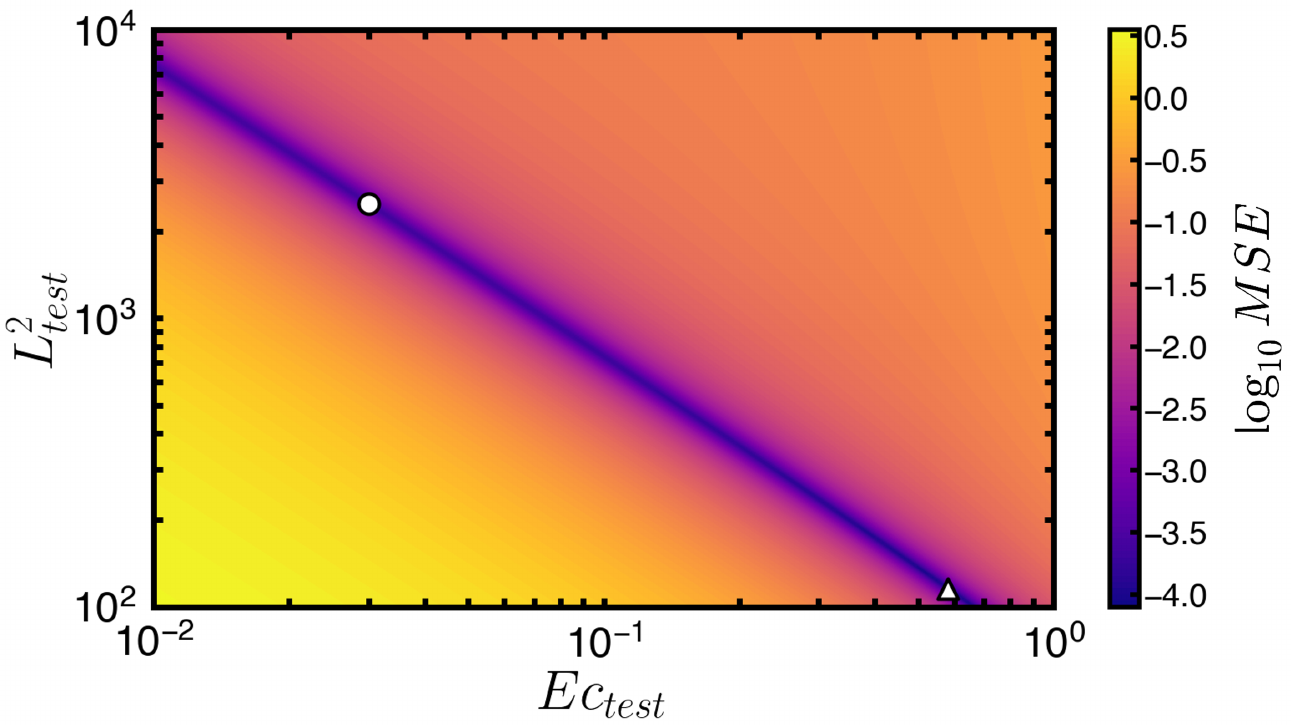}
8I  \caption{Contour plot of the logarithm of the Mean Squared Error ($\log_{10}MSE$) when we fit Eq. (\ref{eq:fenep}) with fixed value $De^{[FENE]}_{fit}=1$ and for a range  $Ec_{test}$ and $L^2_{test}$ values to the data generated by numerical simulations of the FENE-P model for $L^2_{input}=2500$. The colorbar corresponds to the magnitude of $\log_{10}MSE$ for each $(Ec_{test}, L^2_{test})$ combination. The white circle and triangle symbols indicate respectively the known input values $(Ec_{input} = 0.03, L^2_{input}=2500)$, and the optimal fitted values as predicted by Eq. (\ref{eq:fenep}) $(Ec^{[FENE]}_{fit}=0.58, L^2 {}^{[FENE]}_{fit}=115.7)$ for fixed $De^{[FENE]}_{fit}=1$.}
  \label{fig:MSE}
  \end{figure}
 
 To understand in more detail the deviations observed in Table \ref{table_FENE} for the fitted values of $(Ec^{[FENE]}_{fit}, L^2 {}^{[FENE]}_{fit})$ given the known (input) values of $(Ec_{input}, L^2_{input})$, we also compute the solution of Eq. (\ref{eq:fenep}) with fixed $De_{fit}{}^{[FENE]}=1$ and for various elasto-capillary number and finite extensibility combinations, represented by $Ec_{test}$ and $L^2 _{test}$, respectively. To capture the range of values studied by \citet{Wagner2015} as well as the input values ${Ec_{input}, L^2_{input}}$ utilised in this work, we explore a representative range of $10^{-2}\leq Ec_{test}\leq 1$ and $10^2 \leq L^2 _{test} \leq 10^4$. In Figure \ref{fig:MSE} we show (in log scale) a contour plot of the log Mean Squared Error ($\log_{10} MSE$) of the output of Eq. (\ref{eq:fenep}) when it is fitted to test data over this range as a function of $Ec$ and $L^2$.  We compute the error as $\log_{10} MSE = \sum_{i=1}^{n} (\xi(t-t_1)_i - \hat{\xi}(t-t_1)_i)^2/ n$; here, $\xi(t-t_1)_i$ is the actual output of the FENE-P analytical solution for the \textit{i-th} instance when ${Ec_{input}=0.03, L^2 _{input}=2500}$ are considered; $\hat \xi (t-t_1)_i)$ is the predicted value of Eq. (\ref{eq:fenep}) for each test combination $Ec_{test}, L^2 _{test}$ at time $(t-t_1)_i$, and $n$ is the number of available data points. We also indicate in Figure \ref{fig:MSE} the known input combination $(Ec_{input}=0.03, L^2 _{input}=2500)$ with a white circle symbol, and the ``best'" fit (minimum $MSE$) values $(Ec^{[FENE]}_{fit}=0.58, L^2 {}^{[FENE]}_{fit}=115.7)$ with white triangle symbol. This best fit value is obtained from fitting Eq. (\ref{eq:fenep}) with known $De^{[FENE]}{}_{fit}=1$ to the available simulation data obtained with $L^2 _{input}=2500$. Figure \ref{fig:MSE} shows the existence of a broad trough in $Ec, L^2$ space, where the fitting algorithm determines a local minimum in the mean square error.  The strong (anti)-correlation between the values $\{Ec_{test}, L^2 _{test} \}$ that minimise the MSE arises from the functional form of Eq. (\ref{eq:fenep}). Close inspection shows that the terms $Ec$ and $L^2$ appear repeatedly as products. Regression can identify a locally-optimal value of $\{Ec \times L^2 \}$ that minimises the error in fitting $R_{min}(t)$, but uniquely determining optimal individual values of $Ec$ and $L^2$ is more challenging. However, imprecise determination of the best-fit values of $Ec$ and $L^2$ does not corrupt the (enhanced) robustness in the determination of the fluid relaxation time $\tau$ (or equivalently the dimensionless value of $De$) from experimental observations of the filament neck $R_{min}(t)$. Therefore, the improved values of $De_{fit}$ obtained from regressing Eq. (\ref{eq:fenep}) to experimental test data obtained in DoS (as compared to fitting the asymptotic Oldroyd-B result or using the Anna-McKinley form (Eq. (\ref{eq:AM})) provide a promising direction for ensuring more robust extensional rheological measurements of dilute polymeric solutions with Dripping-onto-Substrate rheometry.


\section{Conclusions \label{sec:Conclusions}}

We have studied the thinning and breakup of a liquid droplet that is dripping onto a partially wetting substrate through axisymmetric numerical simulations, using a FENE-P constitutive relation which accounts for finite polymer chain extensibility. We have used the open-source code Basilisk, which is based on a Volume-of-Fluid interface-capturing methodology, and utilises adaptive mesh refinement for accurate and efficient free surface flow solutions. The simulation begins as a prolate drop of fluid is brought into contact with the solid substrate at its base, establishing a macroscopic contact angle as the boundary condition that parameterises the substrate wettability. 
The numerical simulations account for capillary, gravitational, inertial, and viscous forces as well as substrate wettability effects and capture successfully the two main regimes that drive the filament thinning in Dripping-onto-Substrate (DoS) rheometry: (1) the dominance of inertial and capillary forces at early times, which result in the radius of the neck decreasing as a $2/3$ power-law in time and, (2) at later times, the establishment of a characteristic elasto-capillary balance which leads to a period of exponential filament thinning, in which nonlinear fluid elasticity stabilises the thinning process, resulting in thin long-lived viscoelastic ligaments.  

Successful DoS experiments require that the fluid under study exhibits sufficiently large viscoelasticity (i.e. $De = \tau/ \sqrt{\rho R^3 _0 / \gamma} \geq 0.5$) and possess a large enough extensibility of the polymeric chains (i.e. $L^{2} \geq 1600$) for a clear elasto-capillary balance to be established. Moreover, for the first time, the role of gravity (parameterised by a dimensionless Bond number, $Bo = \rho g R_0 \ell_0 /\gamma$), and the wettability of the substrate (parameterised by the macroscopic contact angle $\theta_E$) have both been investigated systematically. Increasing the Bond number and the substrate wettability both are seen to accelerate the initial rate of inertio-capillary necking and the transition to the elasto-capillary balance, without affecting the subsequent evolution of the exponential decrease in filament radius during the elasto-capillary thinning. Our computations show that robust measurements of the polymer relaxation time can be determined in DoS rheometry over a range of conditions but that operational limits of DoS measurements can be optimised by tuning the relative magnitudes of the Bond number and the substrate wettability. 
In addition, we have also developed a simple one-dimensional thinning model which is shown to be capable of capturing the initial Inertio-Capillary (IC) and the subsequent characteristic Elasto-Capillary (EC) thinning regimes. With this model, we can describe the weak but systematic dependence of the inertio-capillary balance on both gravitational body force and the macroscopic contact angle through the variation in the pre-factor $\alpha$ of the inertio-capillary balance that results in Eq. (\ref{eq:IC}). We also report simple power-law correlations of this pre-factor with variations of $Bo$ and wettability (as represented by the factor $cos(\theta_E)$). 
Finally, for fluids with limited extensibilities that do not exhibit very clear exponential thinning regimes, we have proposed a model-fitting framework which is based on the analytical 1D solution for the thinning and breakup of FENE-P filaments. This improves determination of the true relaxation time of an unknown polymer solution through DoS Rheometry (which can be challenging when polymeric solution samples with limited finite extensibilities are involved).

By considering the critical ranges of key dimensionless parameters $(De$, $L^2$, $Bo$, $\theta_E)$ that govern the performance of DoS rheometry using time-resolved free surface numerical simulations, the design of optimal DoS experiments is now possible. Computational Rheometry thus provides a useful digital design tool for enhancing our ability to successfully measure the transient extensional response of low-viscosity complex fluids.

\section*{Conflicts of interest}

There are no conflicts of interest to report.

\section*{Acknowledgements}

This work is supported by the Engineering and Physical Sciences Research Council, United Kingdom, through the EPSRC PREMIERE (EP/T000414/1) Programme Grant. Support from Johnson Matthey for Konstantinos Zinelis is also gratefully acknowledged.

\bibliography{ref} 
\bibliographystyle{rsc} 

\balance
\end{document}


\newpage

\section{Summary}
In the supporting materials, we provide important details on the grid cell resolution of the numerical simulations of a Dripping-onto-Substrate (DoS) experiment \citep{Dinic2015, Dinic2017a, Dinic2019} and the fitting methods implemented for the determination of the relaxation time of a polymer using DoS rheometry. In the first section, we present the mesh convergence study for evaluating the accuracy of the numerical simulations using the code \textit{Basilisk} \citep{Popinet2009} to describe the thinning dynamics of a viscoelastic filament in a DoS experiment. In the second section, we provide a qualitative comparison between the predictions of the ``Anna-McKinley" empirical model \citep{Anna2001} and the FENE-P analytical solution \citep{Wagner2015} in order to determine the apparent relaxation time of a dilute polymer solution. In the last section, we focus on the predictions of the FENE-P analytical solution and we carefully inspect the fitting results for the elastic modulus and the polymer chain extensibility.

\section{Mesh Convergence Study}\label{sec:Mesh}
%
%
\begin{figure}[H]
  \includegraphics[width=0.65\textwidth]{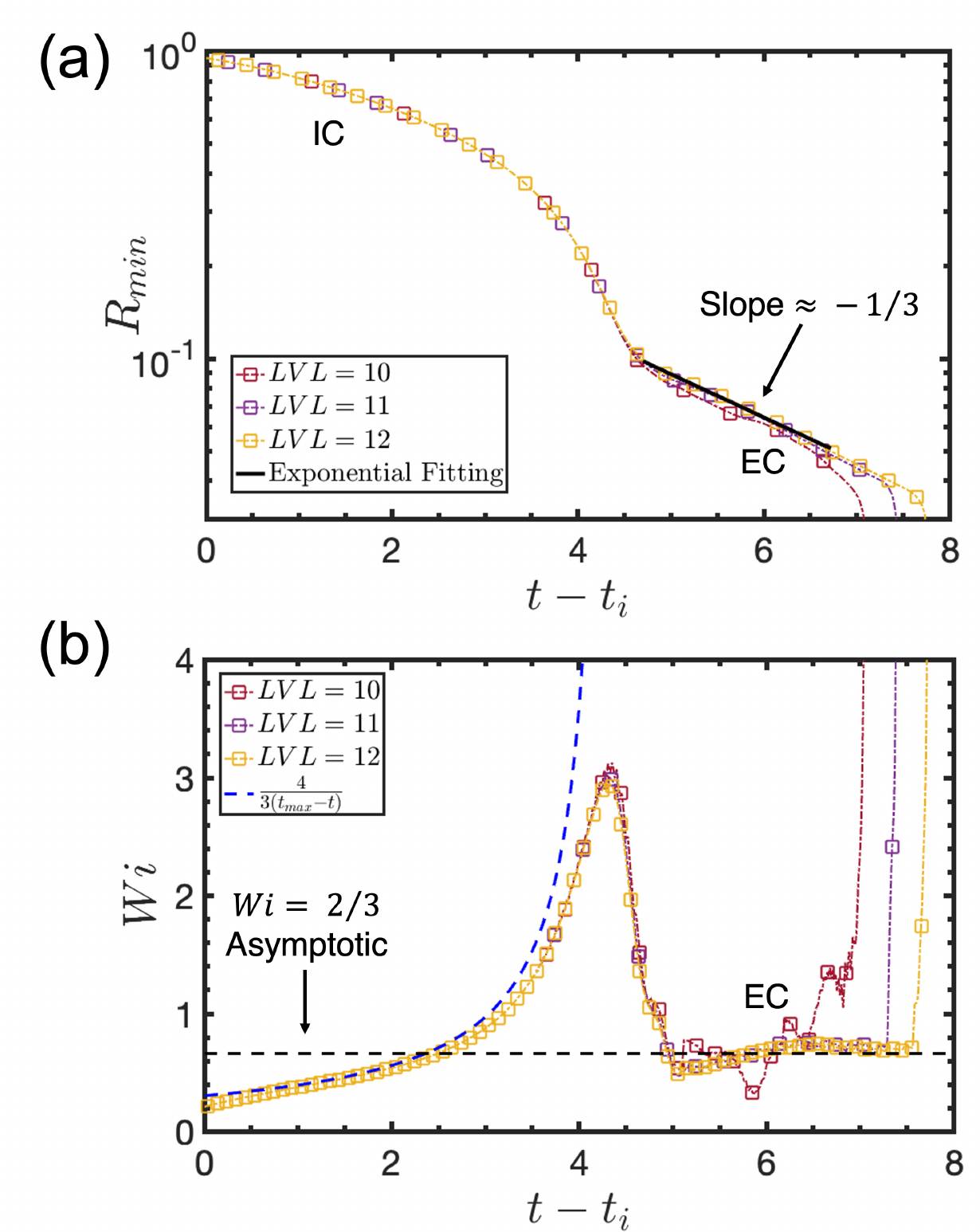}
  \caption{Mesh convergence study using three levels-of-refinement $LVL= 10$, 11 and 12 ($\Delta r_{minimum} = 0.0051$, 0.0026, and 0.0013) for the parameters shown in Table 1 (main text) for the numerical simulations of a thinning filament during a Dripping-onto-Substrate (DoS) experiment; results are presented through the temporal evolution of (a) the minimum filament radius $R_{min}(t)$, and (b) the evolution in the dimensionless strain rate $Wi(t) = \dot{\epsilon}(t) \tau$, in order to evaluate the resolution of the Inertio-Capillary (IC), the subsequent Elasto-Capillary (EC), and finally the Terminal Visco-Elasto-Capillary (TVEC) regimes. The Oldroyd-B asymptotic thinning rate \citep{entov_yarin_1984, Bazileveskii1990, Entov1997, Tirtaatmadja2006}($-1/3$ slope) during the exponential thinning of the filament (black solid line), and the theoretical predictions for the filament dynamics during the IC regime (blue dashed line) are also shown in (a) and (b), respectively. Here, the dimensionless time scale $t$ has been shifted by an initial offset, $t_i$ which corresponds to the time at which the residual effects of the initial spreading process become negligible. The time when the local $Wi$ attains its maximum value is denoted by $t_{max}-t_i = 4.29$ (with $t_{max} = 11.75$ and $t_i=7.46$); this also coincides with the end of the IC regime and the subsequent transition to the EC regime.}
  \label{fig:Mesh}
\end{figure}
We present in Figure \ref{fig:Mesh} the capability of \textit{Basilisk} \citep{Popinet2009} to resolve the filament thinning dynamics during a Dripping-onto-Substrate rheometry adequately, accounting for the wetting process of the substrate. Specifically, we show in Figure \ref{fig:Mesh}(a) and (b) the temporal profiles of the dimensionless minimum radius of the viscoelastic filament $R_{min}(t)$ and the local dimensionless strain rate $Wi(t) = \dot{\epsilon}(t) \tau$ respectively, with $\dot{\epsilon}(t) = -2 d\log(R_{min})/dt$ and $\tau$ denotes the relaxation time of the polymer, for the same physical properties listed in Table 1 (main text) and for three different levels of refinement ($LVL=10$, 11, and 12) which correspond to three distinct minimum square cell sizes ($\Delta r_{minimum} = 0.0051$, 0.0026, and 0.0013). In addition, in the current work, we treat the lateral spreading of the fluid over the solid substrate by setting the macroscopic equilibrium contact angle as a boundary condition on the solid boundary in combination with the height-function method \citep{Afkhami2008}. Although we enforce a no-slip condition on the substrate, the velocity field for the interface advection is located at the centre of the cell faces and therefore \textit{Basilisk} \citep{Popinet2009} allows for an implicit slip condition at the contact line singularity with slip length of $\Delta r_{minimum}/2$ \citep{Afkhami2008, Snoeijer2013}. Hence, these three values of the refinement level (or the minimum cell sizes) also determine the resulting slip lengths at the contact line that can be resolved here (which at $LVL=12$ are up to 4 orders of magnitude lower than the initial filament radius $R_0$).

In Figure \ref{fig:Mesh} we observe that even though the profiles of $R_{min}(t)$ and $Wi(t)$ at $LVL=10$ are seen to fluctuate more compared to $LVL=11$ and 12, all the $LVL$ values exhibit identical Inertio-Capillary (IC) dynamics, and converge to the Oldroyd-B asymptotic of $-1 / (3De)$ \citep{entov_yarin_1984, Bazileveskii1990, Entov1997, Tirtaatmadja2006} (or equivalently $Wi =2 /3$) during the Elasto-Capillary (EC) regime. Moreover, at the highest level of refinement $LVL=12$ (i.e. the lowest $\Delta r_{minimum} = 0.0013$), we do not observe any substantial difference during the exponential EC thinning of the filament. Mesh resolution limitations only become apparent in the terminal linear visco-elasto-capillary (TVEC) thinning at $t-t_i \geq 7$ \citep{Entov1997, McKinley2005, Wagner2005}, which is beyond the scope of the current work. Therefore, at $LVL=11$ ($\Delta r_{minimum}= 0.0026$) we can sufficiently resolve both the initial inertio-capillary thinning and the subsequent onset and evolution of the EC regime, which are the main focus of this work. Finally, we also see that the macroscopic filament dynamics remain essentially unaffected by the magnitude of the slip length ($\sim 10^{-3} R_0$ for $LVL=11$ and $\sim 10^{-4} R_0$ for $LVL=12$) at the contact line singularity.

\section{Fitting results with the Anna-McKinley model and the FENE-P analytical solution}

\begin{figure}[H]
\centering
  \includegraphics[width=0.65\textwidth]{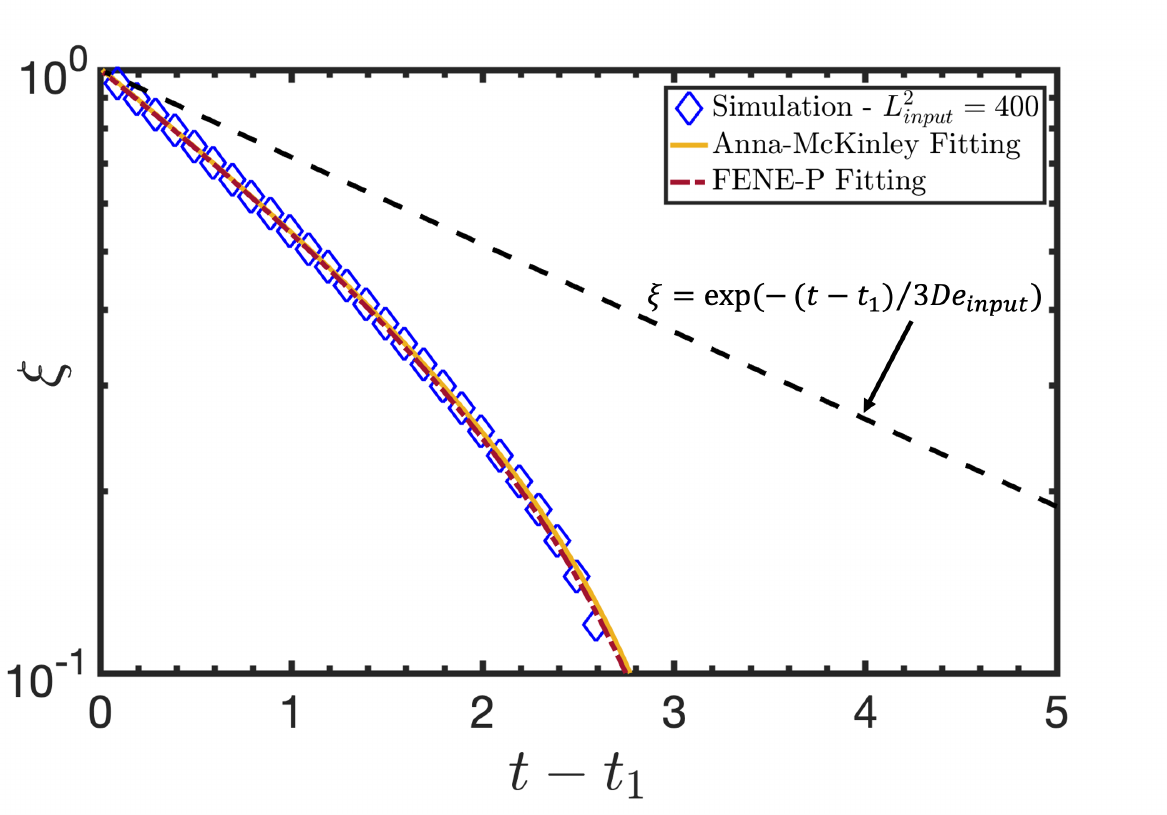}
  \caption{Simulation data of the dimensionless minimum filament radius $\xi$ during the EC regime denoted by diamond symbols for a polymer finite extensibility $L^2_{input} = 400$, fixed Deborah number $De_{input}=\tau/ t_R = 1$, and elasto-capillary number $Ec_{input}=\eta_p R_0/(\tau \gamma) = 0.03$. Here, the time $t$ is shifted by $t_1$ which corresponds to the onset of the EC regime. The predictions of the Anna-McKinley fitting model (Eq. (18) in the main text) and the FENE-P analytical solution (Eq.(19) in the main text) for $L^2_{input} = 400$ are shown by solid orange and dashed red lines, respectively. The dashed black line corresponds to the prediction of exponential Oldroyd-B decay during the EC regime \citep{Entov1997, Clasen2006, Rajesh2022}.}
  \label{fig:Fitting}
\end{figure}

In Figure \ref{fig:Fitting} we show the results obtained from fitting Eq. (18) labelled ``Anna-Mckinley Fitting" (solid orange line) and Eq. (19) labelled ``FENE-P Fitting" (dashed red line) to the data (blue diamonds) generated with the numerical simulations presented in this work for the evolution of the scaled filament radius $\xi = R_{min}(t)/ R_1$ with time $t-t_1$, where $R_1$ and $t_1$ are the dimensionless filament radius and time at the onset of the EC regime, respectively. The properties considered here are the same as in Table 1 (main text) but with the smallest finite extensibility of the polymer ($L^2 = 400$). We observe a significant deviation of the simulation and the predictions of the two fitting equations from the Oldroyd-B asymptotic result (dashed black line) during the EC regime. We also see that both Eq. (18) and (19) overlap very well with the simulation data qualitatively. However, Table 6 and Figure 12 (main text) reveal important quantitative differences that should be carefully considered when we aim to extract the apparent relaxation time of dilute polymer solutions with limited extensibility.

\section{Analysis of the predictions of the FENE-P analytical solution \label{sec:MSE}}

\begin{figure}[H]
\centering
  \includegraphics[width=0.65\textwidth]{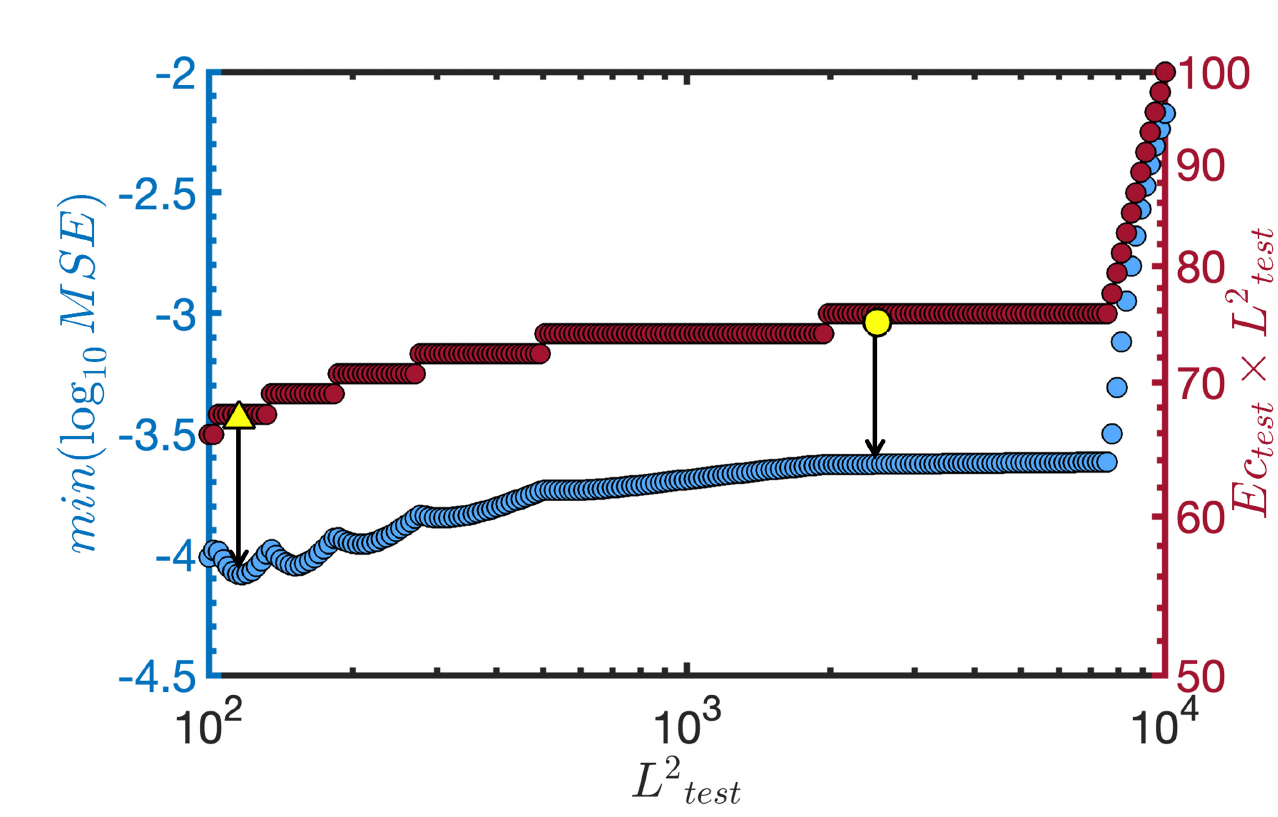}
  \caption{Minimum Mean Squared Error ($min(\log_{10} MSE)$) in blue and the combinations of the elasto-capillary number and polymer finite extensibility ($Ec_{test} \times L^2 _{test}$). The values correspond to the minimum Mean Squared Error from the predictions of the FENE-P analytical solution for filament thinning during the EC regime for different values of polymer extensibility ($L^2 _{test}$). The known input and the predicted values of the products ($Ec_{test} \times L^2 _{test}$) are denoted by a yellow circle and triangle symbol, respectively. The black arrows point to the minimum Mean Squared Error, $min(\log_{10} MSE) \approx -3.7$ and  $min(\log_{10} MSE) \approx -4$, of the known and predicted values ($Ec_{test} \times L^2 _{test}$), respectively.}
  \label{fig:MSE}
\end{figure}

We present in Figure \ref{fig:MSE} the variation in the minimum Mean Squared Error (MSE), and the product ($Ec_{test} \times L^2 _{test}$) of the different combinations of elasto-capillary numbers (which correspond to the dimensionless elastic modulus, $Ec = G R_0/ \gamma$) and polymer extensibilities that result in the minimum MSE ($min(\log_{10} MSE)$). The error evolves continuously with the value of finite extensibilities $L^2 _{test}$ ($10^2 \leq L^2_{test} \leq 10^4$) considered in Figure 14 (main text). 
The product $Ec_{test} \times L^2_{test} =75$ of the known (ground-truth) values ($Ec_{test}= 0.03, L^2_{test} =2500$) is denoted by a yellow circle symbol in Figure \ref{fig:MSE}, and results in a minimum MSE of $min(\log_{10} MSE)\approx -3.7$. These are the elasto-capillary number ($Ec_{test}= 0.03$) and finite extensibility value ($L^2_{test} =2500$) that we expect from the predictions of Eq. (19) (main text). However, the optimiser of the fitting process using Eq. (19) detects a different value of the minimum MSE. This value is $min(\log_{10} MSE) \approx -4$ and corresponds to the product $Ec_{test} \times L^2_{test} \approx 67.1$ ($Ec_{test}= 0.58, L^2_{test} =115.7$) denoted by a yellow triangle symbol in Figure \ref{fig:MSE}. Nonetheless, this systematic error in the prediction of the elasto-capillary number and polymer chain extensibility does not substantially affect the determination of the relaxation time of the polymer using the FENE-P analytical solution (Eq. (19) in the main text).


\bibliography{ref}